\documentclass[11pt,a4paper]{article}
\pdfoutput=1
\usepackage{jheppub}
\usepackage{psfrag}
\usepackage{slashed}
\usepackage{cancel}
\usepackage{lscape}

\usepackage{caption}
\usepackage{array}
\usepackage{graphicx}
\usepackage{subcaption}

\definecolor{mygreen}{rgb}{0,0.7,0}

\def\la{\langle}
\def\ra{\rangle}
\def\A#1#2{\la#1#2\ra}
\def\B#1#2{[#1#2]}
\def\AB#1#2#3{\la#1|#2|#3]}

\DeclareMathOperator{\tr}{\rm tr}
\def\trm{\tr_-}
\def\trp{\tr_+}
\def\MP#1#2{(#1\cdot#2)}
\def\trfive{\tr_5}

\def\eps{\epsilon}
\def\fl#1{#1^\flat}

\def\cv#1#2{\AB{#1}{\gamma^\mu}{#2}}

\def\fl#1{{#1^{\flat}}}
\def\flm#1{{#1^{\flat,\mu}}}

\newcommand{\tekst}{\textrm}
\newcommand{\kur}[1]{\mathcal{#1}}
\newcommand{\nn}{\nonumber}
\newcommand{\feyn}[1]{#1\kern-0.45em/}
\newcommand{\id}{\mathrm{d}}
\definecolor{myblue}{rgb}{0,0,0.7}

\definecolor{Carnelian}{rgb}{0.701961, 0.105882, 0.105882}

\preprint{}

\title{A Two-Loop Five-Gluon Helicity Amplitude in QCD}

\author[a]{Simon Badger}
\author[a,b]{Hjalte Frellesvig}
\author[a]{Yang Zhang}
\affiliation[a]{
Niels Bohr International Academy and Discovery Center, The Niels Bohr Institute,\\%
University of Copenhagen, Blegdamsvej 17, DK-2100 Copenhagen, Denmark}
\affiliation[b]{
Istituto Nazionale di Fisica Nucleare, Sezione di Roma,
P.le Aldo Moro 2, 00185 Roma, Italy
}
\emailAdd{badger@nbi.dk,hjf@nbi.dk,zhang@nbi.dk}

\abstract{%
We compute the planar part of the two-loop five gluon amplitude with all helicities positive. To
perform the calculation we develop a $D$-dimensional generalized unitarity procedure allowing us
to reconstruct the amplitude by cutting into products of six-dimensional trees. We find a compact
form for the integrand which only requires topologies with six or more propagators. We perform
cross checks of the universal infra-red structure using numerical integration techniques.
}
\notoc

\begin{document}
\maketitle
\flushbottom

\section{Introduction}

Precision QCD looks set to play a leading role in the next phase of LHC operations and forthcoming
analyses. Next-to-leading order (NLO) and in some cases next-to-next-to-leading order (NNLO) are
extremely important for accurate modelling of QCD backgrounds to new physics searches as well as
measurements of standard model parameters such as the strong coupling, $\alpha_s$. Predictions at
this level of accuracy require complicated loop amplitude computations which have always been a
bottleneck using the traditional Feynman diagram approach.  The number of terms appearing in
intermediate expressions grows extremely fast with additional external particles and especially when
increasing the loop order. Unitarity and on-shell methods
\cite{Bern:1994zx,Bern:1994cg,Britto:2004ap} have been developed to allow computations using only
the physical degrees of freedom, reducing the complexity. The principle of the generalized unitarity
method \cite{Britto:2004nc,Ellis:2007br,Forde:2007mi,Giele:2008ve,Badger:2008cm} is to reduce the loop amplitudes
to the sum over multiple cuts where each cut factorizes into the product of tree-level amplitudes.
The integrand reduction method of Ossola, Papadopoulos and Pittau (OPP)~\cite{Ossola:2006us} shows
how to systematically remove singularities from previously computed cuts so a systematic, top-down
approach can be taken. The complete method yields a completely algebraic approach to the computation
of one-loop amplitudes and has been automated in several numerical algorithms
\cite{Ossola:2007ax,Berger:2008sj,Giele:2008bc,Ellis:2008qc,Mastrolia:2010nb,Badger:2010nx,Hirschi:2011pa,Bevilacqua:2011xh,Cullen:2011xs,Badger:2012pg}.

Multi-loop methods for highly symmetric theories such as $\mathcal{N}=4$ super Yang-Mills theory are
by now extremely advanced~\cite{Bern:1997nh,Bern:2005iz,Bern:2006ew,Bern:2007ct,ArkaniHamed:2010kv}.
Similar progress in multi-leg QCD computations has however not been possible, mainly due to the much
larger set of master integrals appearing in the amplitudes. State of the art computations in QCD
have been completed for most $2\to2$ scattering processes where a Feynman diagram approach combined
with integration-by-parts (IBP) identities~\cite{Chetyrkin:1981qh} has been successful
~\cite{Anastasiou:2000kg,Anastasiou:2000ue,Anastasiou:2001sv,Glover:2001af,Garland:2001tf,Garland:2002ak,Gehrmann:2011aa}.
Unitarity based methods have also played an important role in computations of a similar level of
complexity \cite{Bern:2000dn,Bern:2000ie,Bern:2001df,Bern:2001dg,Bern:2002tk,Bern:2003ck}. Recently,
the first genuine NNLO QCD corrections to $2\to2$ scattering have been computed, after cancelling
the infra-red divergences and performing the phase-space integration
\cite{Czakon:2013goa,Boughezal:2013uia,Ridder:2013mf}.

Motivated by the successes of one-loop techniques, there has been recent progress in extending
generalized unitarity and integrand reduction methods for applications in multi-leg two-loop
amplitudes. The IBP identities have been understood in a unitarity compatible form which has shed
light on the integral basis \cite{Gluza:2010ws}. Following this direction, the maximal unitarity
method has been developed for maximal cuts of two-loop massless and massive amplitudes
\cite{Kosower:2011ty,Larsen:2012sx,Johansson:2012zv,CaronHuot:2012ab,Johansson:2013sda,Sogaard:2013yga}.
Integrand reduction methods have also been developed to the two-loop level \cite{Mastrolia:2011pr,
Badger:2012dp} via the polynomial fitting and Gram matrix constraints. Furthermore, the integrand
reduction method was systematically generalized to all loop-orders using
computational algebraic geometry \cite{Zhang:2012ce,Mastrolia:2012an}.
This approach has been applied to a number of two-loop
\cite{Mastrolia:2012wf,Mastrolia:2013kca} and three-loop \cite{Badger:2012dv} examples.
These approaches offer the benefit that they apply to arbitrary gauge theories rather than being limited
to super-symmetric amplitudes. Understanding the role of algebraic geometry in these methods has
been particularly important and some of the more formal mathematical aspects have also been recently explored
\cite{Huang:2013kh}.

The aim of this paper is to generalize the integrand reduction method to dimensionally regulated
amplitudes in a way compatible with generalized unitarity cuts.
Though the planar part of the five-gluon amplitude with all positive helicities formally contributes at N$^{3}$LO
it is, to the best of our knowledge, the first computation of a five-point amplitude in a non-super-symmetric theory.
In order to perform the multiple cuts in $D$ dimensions it is necessary to
consider tree-level amplitudes in minimum six dimensions. We make use of the six-dimensional
spinor-helicity formalism \cite{Cheung:2009dc} which has been used previously for generalized cuts at
one-loop \cite{Bern:2010qa,Davies:2011vt}. The five-scale kinematics algebra can be treated
efficiently when written in terms of momentum twistors \cite{Hodges:2009hk} enabling the final
result to be written in a particularly compact form which we present in eq. \eqref{eq:primitive}.

Our paper is organized as follows: In section \ref{sec:ddreduction} we outline the generalization of
the integrand reduction methods to $D$ dimensions. We prove that the approach is compatible with the
fitting of each integrand from the product of six-dimensional tree-level amplitudes via generalized
unitarity cuts.  As the first non-trivial application of the integrand reduction procedure at
two-loops, we present the planar five-gluon amplitude with all positive helicities in section
\ref{sec:5gresult}. We present a numerical evaluation of the amplitude in Section \ref{sec:5gnum}
and check the universal infra-red properties before presenting our conclusions. We include an
appendix describing the explicit parametrization of the kinematics in terms of momentum twistors
used to simplify the computation, and one listing the Feynman rules we use for tree-level
calculations.

\subsection{Notation}

The paper will adopt a fairly conventional approach to the spinor products and
Lorentz products, nevertheless we outline them here for clarity. External momenta
will be denoted $p_i^\mu$ with the usual short-hand notation for their sums and invariants,
\begin{align}
p_{ij} &= p_i + p_j, & s_{ij} &= p_{ij}^2.
\end{align}
Spinor products are constructed from holomorphic ($\lambda_\alpha$) and anti-holomorphic ($\tilde{\lambda}_{\dot{\alpha}}$)
two-component Weyl-spinors
\begin{align}
  \A{i}{j} &= \lambda_{\alpha}(p_i)\lambda^{\alpha}(p_j), &
  \B{i}{j} &= \tilde{\lambda}^{\dot{\alpha}}(p_i)\tilde{\lambda}_{\dot{\alpha}}(p_j),
\end{align}
such that $\A{i}{j}\B{j}{i} = s_{ij}$. We find that the amplitudes are conveniently written in terms of traces over \(\gamma\)-matrices:
\begin{align}
\tr_{\pm}(abcd) &= \frac{1}{2} \tr \! \big( (1 \pm \gamma_5) \feyn{p}_a \feyn{p}_b \feyn{p}_c \feyn{p}_d \big),
\end{align}
where the parity odd contracted anti-symmetric tensor $\tr_5 =
4i\varepsilon_{\mu_1\mu_2\mu_3\mu_4} p_1^{\mu_1} p_2^{\mu_2} p_3^{\mu_3} p_4^{\mu_4}$ is constructed
by the linear combination,
\begin{align}
\trfive &= \tr \! \left( \gamma_5 \feyn{p}_1 \feyn{p}_2 \feyn{p}_3 \feyn{p}_4 \right) \; = \; \B12 \A23 \B34 \A41 - \A12 \B23 \A34 \B41 .
\label{trfivedef}
\end{align}

\begin{figure}[h]
  \begin{center}
    \includegraphics[width=0.5\textwidth]{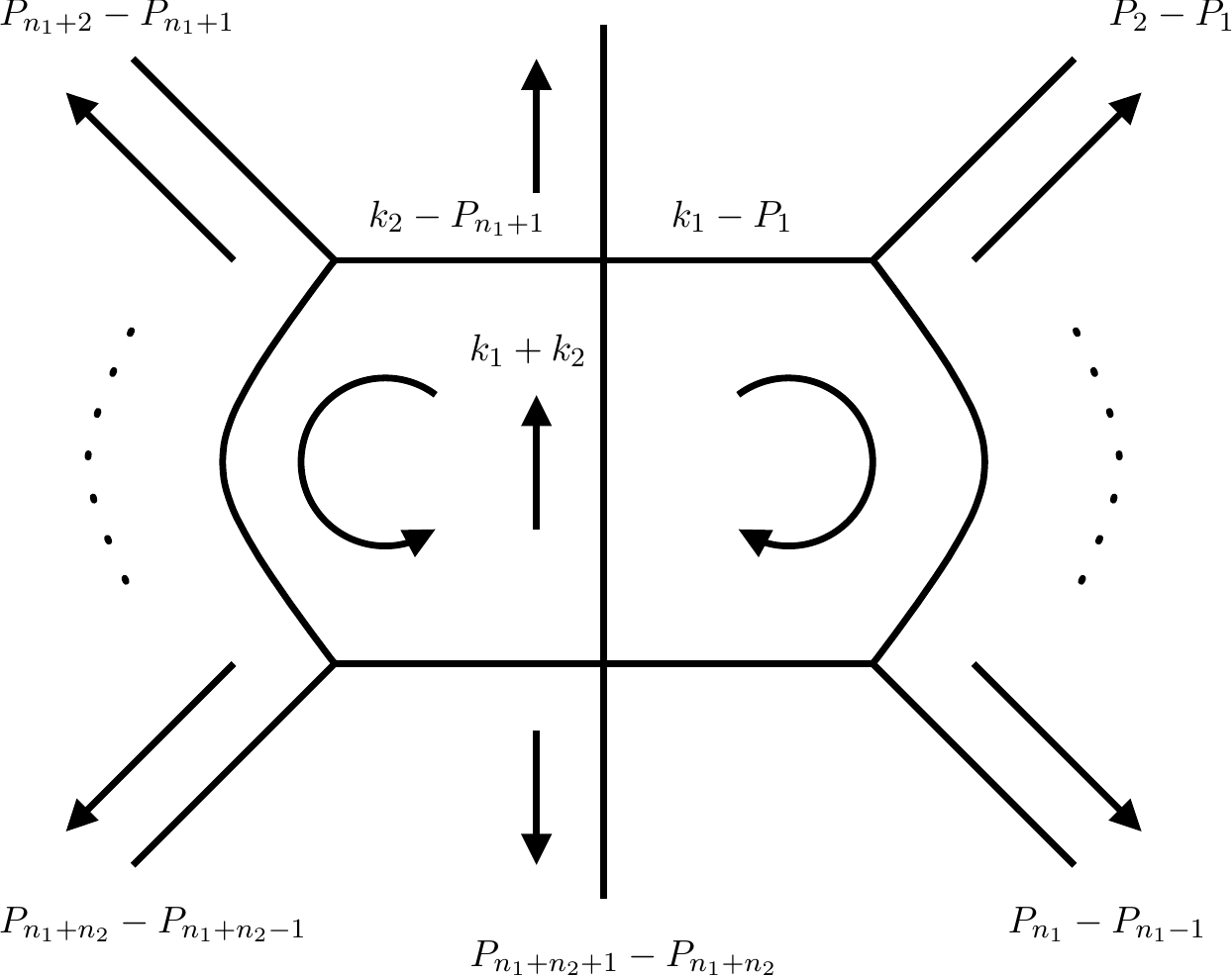}
  \end{center}
  \caption{A pictorial representation of the planar two-loop topology denoted $I^{[D]}_{n_1 n_2 1;\kur{P}}$}
  \label{fig:2lnot}
\end{figure}

Most of the calculations in this paper are done in dimensional regularization, with several different dimensions in play simultaneously. They are:
\begin{itemize}
\item[$D\,\,$] The number of dimensions in dimensional regularization. $D=4-2 \epsilon$.
\item[$D_s$] The number of dimensions in which we allow the polarizations directions of internal gluons. In the FDH-scheme (see section \ref{sec:integrandreduction}) $D_s=4$.
\item[$\kur{D}\,\,$] The number of dimensions in which we embed the $D$-dimensional momenta. We will use $\kur{D}=6$.
\end{itemize}
\noindent
The two-loop integrals (and integrands) appearing in the paper will be written as:
\begin{align}
  I^{[D]}&_{n_1 n_2 n_{12}; \kur{P}}[N] = \nonumber\\& \int \frac{\id^D k_1}{(2\pi)^D} \frac{\id^D k_2}{(2\pi)^D}
    \frac{N}{
    \prod\limits_{i=1}^{n_1} (k_1-P_i)^2
    \prod\limits_{j=1+n_1}^{n_1+n_2}(k_2-P_j)^2
    \prod\limits_{h=1+n_1+n_2}^{n_1+n_2+n_{12}} (k_1+k_2-P_h)^2
    },
  \label{eq:2lintnotation}
\end{align}
where \(\kur{P}\) contains any additional information necessary to specify the topology, namely the
configuration of external momenta flowing along each propagator which defines the set $\{P_i\}$. We will
specify a shorthand for $\kur{P}$ on a case by case basis rather than opting for a more general notation. Topologies for
which \(n_{12}=0\) will be referred to as butterfly-type topologies. Figure \ref{fig:2lnot} gives a
pictorial representation of the planar topologies considered in the rest of the paper.

\section{Integrand reduction and generalized unitarity in $D$ dimensions \label{sec:ddreduction}}

In this section we develop a multi-loop integrand reduction procedure valid in $D$ dimensions.
The main aim is to obtain a formalism that allows the integrand form of the amplitude to be computed
from the product of tree-level amplitudes by applying generalized unitarity cuts. In section
\ref{sec:integrandreduction} we will review integrand reduction for multi-loop amplitudes and
emphasize the special features which appear when applying the method to dimensionally regulated
amplitudes. The extra dimensional parts of the amplitude can be expressed using three mass-like
parameters which can be effectively embedded in $\kur{D}=6$ dimensions, as described in section
\ref{sec:scalars}. In order to do calculations in the six-dimensional space, the
six-dimensional spinor-helicity formalism developed by Cheung and O'Connell \cite{Cheung:2009dc},
which is described in section \ref{sec:sixd}, proves itself useful. In section
\ref{sec:unitarity} we will describe how to do generalized unitarity cuts in $D$ dimensions in
the context of our specific parametrization, illustrated by a specific \(2 \rightarrow 2\) example.
In section \ref{sec:feynman} we will comment on how to reproduce the six-dimensional set-up from Feynman diagrams.

\subsection{$D$-dimensional integrand reduction \label{sec:integrandreduction}}

A generic loop diagram (with $L$ loops and $P$ propagators) can be written as
\begin{align}
I &= \int \frac{\id^D k_1}{(2\pi)^D} \ldots \frac{\id^D k_L}{(2\pi)^D} \frac{N}{D_1 \ldots D_P}.
\end{align}
The numerator function \(N\) can be a function of the loop momenta $k_i$ only through scalar
products of the form \(k_i \! \cdot \! k_j\), \(k_i \! \cdot \! p_j\), or \(k_i \! \cdot \!
\omega_j\), where \(p_i\) are the momenta of the external particles, and \(\omega_i\) are vectors
constructed to be perpendicular to all the \(p_i\).  Some of the scalar products can be expressed in
terms of the propagators \(D_i\), giving
\begin{align}
N &= \Delta + \sum_{i=1}^P \kappa_i D_i,
\end{align}
where \(\Delta\) can be expressed polynomially in terms of the remaining scalar products
\((x_1,\ldots,x_n)\), known as irreducible scalar products (ISPs):
\begin{align}
\Delta &= \sum_{i_1 \ldots i_n} c_{i_1 \ldots i_n} x_1^{i_1} \cdots x_n^{i_n}.
\label{cexpansion}
\end{align}
The upper limits in the sum are determined by a division over the Gr\"obner basis, as explained below.

It is required that the reduction to the irreducible numerator \(\Delta\) is maximal, i.e. if
$N$ can be written as a combination of $D_i$'s, then $\Delta$ must be zero. This process
is known as integrand reduction, and the result will be an amplitude split up into a set of
topologies, each characterized by a set of propagators and a corresponding irreducible numerator.

Explicitly, the integrand reduction is achieved by the {\it Gr\"obner
  basis} method and {\it synthetic polynomial
  division}. \cite{Zhang:2012ce, Mastrolia:2012an} (For
mathematical details, see Chapter 2 of \cite{MR2290010}.) The
denominators $D_1$, $\ldots$, $D_P$ generate an {\it ideal},
\begin{equation}
  \label{eq:8}
  I=\langle D_1 ,\ldots, D_P\rangle,
\end{equation}
and we can calculate the Gr\"obner basis $G(I)=\{g_1, \ldots, g_m\}$ of $I$ using a
monomial ordering. Then the polynomial division over $G(I)$ is performed,
\begin{equation}
  N=\Delta + \sum_{i=1}^P q_i g_i,
\label{division}
\end{equation}
where
$\sum_{i=1}^k q_i g_i\in
I$ contributes to diagrams with fewer propagators.  The Gr\"obner basis
ensures that this reduction is maximal. This algebraic geometry
approach works for any number of loops, and both in an integer and a dimensionally regulated number of dimensions.

In practice, if the explicit form of $N$ is known from
Feynman rules, then the division in eq. (\ref{division}) over $G(I)$ directly determines
$\Delta$. Alternatively, we can fit coefficients in eq. (\ref{cexpansion}) by the {\it generalized unitarity
method}, which puts all the propagating momenta on shell,
\begin{equation}
  D_1=\ldots =D_P=0,
\label{cut-equation}
\end{equation}
imposing a set of constraints on the loop momenta \(k_i\). The
solution to the equation system (\ref{cut-equation}) may have
several branches. Mathematically eq. (\ref{cut-equation}) defines an algebraic
set, which decomposes as the union of several {\it affine
  varieties}. In that case the ideal $I$ decomposes as the
intersection of several {\it primary ideals} \cite{MR0463157},
\begin{equation}
  I=I_1 \cap \ldots \cap I_n,
\end{equation}
where each $I_j$ is a primary ideal corresponding to one branch
of the solution. For each branch, the freedom remaining after imposing the unitarity cut constraints can be parametrized by a set of parameters \(\tau_i\), giving
\begin{align}
\Delta|_{\tekst{cut}} &= \sum_{j_1 \ldots j_m} d_{j_1 \ldots j_m} \tau_1^{j_1} \cdots \tau_m^{j_m},
\label{dexpansion}
\end{align}
where \(\Delta|_{\tekst{cut}}\) can be found as a product of tree amplitudes (see section \ref{sec:unitarity}).

Inserting the constrained loop-momenta into eq. (\ref{cexpansion}) allows us to set up a linear relation between the two kinds of coefficients,
\begin{align}
{\bold d} &= M {\bold{c}},
\label{dismc}
\end{align}
with \({\bold c}\) and \({\bold d}\) being vectors of the coefficients
from eq. (\ref{cexpansion}) and
(\ref{dexpansion}) respectively. Solving eq. (\ref{dismc}) for \({\bold c}\) allows us to determine the
irreducible numerator straight from unitarity cuts.

One subtle question is whether eq. (\ref{dismc}) has a unique solution. More explicitly, {\it is
there any polynomial $f$ such that $f$ vanishes at the unitarity cut, but $f\not \in I$?} In that case,
the term $c f$ in the numerator contributes to the integrand basis. However, since $c f$
vanishes at the unitarity cut, the value of $c$ cannot be fixed by polynomial fitting and the solution
of eq. (\ref{dismc}) is not unique.

We show that this problem can be avoided if the ideal $I$ is {\it radical}. The radical of $I$ is
defined as the ideal,
\begin{equation}
  \sqrt I=\{f |f^n\in I, n \in \mathbb N\},
\end{equation}
where $I$ will be a subset of $\sqrt I$. If $I=\sqrt I$, then we
say that $I$ is radical.  Hilbert's Nullstellensatz \cite{MR0463157} states that
$\sqrt I$ is the set of all polynomials vanishing on the cut, and
hence if $I$ is radical, then all the coefficients of
the integrand basis $\Delta$ can be extracted from unitarity cuts.

In this paper we focus on two-loop $D$-dimensional integrand reduction. Specifically we will be
using the four-dimensional helicity scheme (FDH) which consists of leaving the external particles
and all polarizations in four dimensions, but shifting the loop-momenta to \((D = 4-2 \epsilon)\)
dimensions \cite{Bern:2002zk}.

We will handle the $D$-dimensional loop-momenta by splitting them into four-dimensional and
higher dimensional components:
\begin{eqnarray}
  \label{eq:5}
  k_i = \bar{k}_i + k_i^{[-2\epsilon]}, \quad i=1,2 .
\end{eqnarray}
By the symmetry of the higher-dimensional space, the amplitudes can depend on $k_i^{[-2\epsilon]}$
only through the three scalar products,
\begin{equation}
  \label{mudefs}
  \mu_{11} = -(k_1^{[-2\epsilon]} \cdot k_1^{[-2\epsilon]}) \; , \quad
  \mu_{22} = -(k_2^{[-2\epsilon]} \cdot k_2^{[-2\epsilon]}) \; , \quad
  \mu_{12} = -2 (k_1^{[-2\epsilon]} \cdot k_2^{[-2\epsilon]}) \, .
\end{equation}

The $D$-dimensional integrand reduction has several good properties:
\begin{itemize}
\item The ideal $I$ is radical, so all coefficients in the integrand basis can be fixed.
This can be proved as follows: At two loop order there are two types of ISPs, namely $m$ ISPs $\{x_1, \ldots
x_m\}$ of the form $k_i \cdot p_j$ or $k_i \cdot \omega_j$, and the remaining three $\mu_{11}$,
$\mu_{12}$, and $\mu_{22}$. For a diagram with a $(k_1+k_2)$ internal
leg and $P$ propagators, cut equations can be rewritten as the three
quadratic equations $k_1^2=k_2^2=(k_1+k_2)^2=0$ and $P-3$ linear
equations. These linear equations determine $(P-3)$ reducible scalar
products (RSPs), which always have the form $k_i \cdot p_j$. So $m=8-(P-3)=11-P$. 
After eliminating these RSPs, we get the ideal of 
cut equations:
\begin{equation}
  \label{eq:9}
 I=\langle \mu_{11}-f_1(x_1,\ldots x_m),\quad \mu_{12}-f_2(x_1,\ldots
  x_m),\quad \mu_{22}-f_3(x_1,\ldots x_m)\rangle.
\end{equation}
We then have the following map,
\begin{equation}
  \label{eq:10}
 \phi: \mathbb C[x_1,\ldots x_n, \mu_{11},\mu_{12}, \mu_{22}]/I \to
  \mathbb C[x_1,\ldots x_n] ,
\end{equation}
with $\mu_{11} \mapsto f_1(x_1,\ldots x_m)$, $\mu_{12} \to
f_2(x_1,\ldots x_m)$ and $\mu_{22} \to f_3(x_1,\ldots x_m)$.
It is clear that $\phi$ is an isomorphism, and since $\mathbb C[x_1,\ldots x_n]$
is a domain, $I$ is a prime ideal.
A prime ideal must be radical, which proves the
proposition. Similarly, for a butterfly diagram without any $(k_1+k_2)$ internal
leg, $I=\langle \mu_{11}-f_1(x_1,\ldots x_m),
\mu_{22}-f_3(x_1,\ldots x_m)\rangle$, $m=10-P$ and the proof is similar.

\item Since $I$ is prime, the primary decomposition is trivial and there is only one
  branch of the unitarity cut.

\item The unitary cut solution always has $11-P$ degrees of freedom.
For a diagram with a $(k_1+k_2)$ internal
leg and $P$ propagators, $m=11-P$, so by the isomorphism $\phi$,
 \begin{equation}
    \label{eq:doubleboxdim}
    \dim \mathcal Z(I)=\dim \mathbb C[x_1,\ldots x_m,
    \mu_{11},\mu_{12}, \mu_{22}]/I =\dim
    \mathbb C[x_1,\ldots x_m]=11-P,
  \end{equation}
  where $\mathcal Z(I)$ is the \textit{zero locus} of $I$~\cite{MR0463157}.
Similarly, for a butterfly diagram with $P$ propagators, $m=10-P$, so
 \begin{equation}
    \label{eq:butterflydim}
    \dim \mathcal Z(I) =\dim
    \mathbb C[x_1,\ldots x_m, \mu_{12}]=m+1=11-P.
  \end{equation}
This conclusion implies that a diagram and its parent diagram must
have different cut solutions, since the solutions have different
dimensions. In the four-dimensional case, there are examples in which a diagram
and its parent diagram have the same cut solutions, and it is then
difficult to carry out the subtraction, however, for cuts in $D=4-2\epsilon$
dimensions, this difficulty is avoided.

\end{itemize}

\subsection{Tree-level amplitudes with the six-dimensional spinor-helicity formalism \label{sec:scalars}}

From eq. (\ref{mudefs}), we saw that the set of loop-momenta gets three extra components $\mu_{11}$, $\mu_{22}$, and $\mu_{12}$ in dimensional reduction, and to embed those we need at least a six-dimensional space. A specific embedding in six dimensions is
\begin{align}
K_1 &= \big( k_1, m_1 \cos \theta_1 , m_1 \sin \theta_1  \big) \,, & K_2 &= \big( k_2, m_2 \cos \theta_2 , m_2 \sin \theta_2  \big) \,,
\end{align}
with \(m_1^2 = \mu_{11}\), \(m_2^2 = \mu_{22}\), and the angles being related by
\begin{align}
\cos( \theta_2 - \theta_1 ) = \frac{\mu_{12}}{2 m_1 m_2}.
\end{align}
The fourth degree of freedom is left free, but due to the Lorentz symmetry of the extra dimensions final results will be independent of its value.

Thus we need to calculate a numerator with the loop-momenta living in \(\mathcal{D} \geq 6\) dimensions but with
the polarizations of gluons circulating in the loops living in \(D_s=4\) dimensions. We will now for a
moment treat \(D_s\) as a free parameter. If \(D_s>\mathcal{D}\), a careful consideration
\cite{Bern:2002zk,Giele:2008ve} shows that each of the \((D_s-\mathcal{D})\) higher dimensional components
will act as a scalar-like particle, which behaves just like the coloured scalars known from
\(\kur{N}=4\) super Yang-Mills. This means that in addition to the three-point gluon-scalar-scalar-interaction,
it has four-point vertices with gluons and with other scalars. All the Feynman rules for gluons and
scalars are listed in appendix \ref{feynmanappendix}.

Expressed in terms of the scalar particle, we get that for pure Yang-Mills theory
\begin{align}
  \Delta_{g}^{[D_s]} &=
  \Delta_g^{[\mathcal{D}]} +
  (D_s-\mathcal{D}) \Delta_{s}^{[\mathcal{D}]} +
  (D_s-\mathcal{D})^2 \Delta_{2s}^{[\mathcal{D}]},
\label{scalareqn}
\end{align}
where \(\Delta_{s}\) and \(\Delta_{2s}\) are the contributions from diagrams with respectively one and two scalar loops.
We will perform the computation using $\mathcal{D}=6$ and analytically continue eq.~\eqref{scalareqn} to $D_s < 6$.
We note that the FDH scheme described in the previous section, corresponds to taking $D_s\to4, D\to4-2\eps$,
but alternatively one can chose $D_s = D \to 4-2\eps$ which corresponds to the 't Hooft-Veltman scheme \cite{Bern:2002zk}.

\subsection{The six-dimensional spinor-helicity formalism \label{sec:sixd}}

To calculate the six-dimensional numerators, we will be dealing with six-dimensional momenta and
polarizations, and a convenient way to handle those is the six-dimensional spinor-helicity formalism
developed by Cheung and O'Connell \cite{Cheung:2009dc}.  In six dimensions, Weyl-spinors are defined
as \(\Lambda^{Aa}\) and \(\tilde{\Lambda}_A^{\dot{a}}\), with \(A\) being a spinor-index running
from \(1\) to \(4\), and \(a\) and \(\dot{a}\), each running from \(1\) to \(2\), being indices of
the little group.

Adopting a notation where \(\Lambda\) is denoted by an angle bracket and \(\tilde{\Lambda}\) by a square bracket
\begin{align}
\Lambda^{Aa} &= \langle p^a | \; , & \tilde\Lambda_{A \dot{a}} &= [ p_{\dot{a}} | \; ,
\end{align}
the six-dimensional spinors obey a set of relations similar to Weyl-spinors in four dimensions,
\begin{align}
\Lambda^{Aa} \Lambda_a^{B} &= P_{\mu} \tilde{\Sigma}^{\mu AB} \,, &
\tilde{\Lambda}_{A \dot{a}} \tilde{\Lambda}_{B}^{\dot{a}} &= P_{\mu} \Sigma_{AB}^{\mu} \,,
\end{align}
with $ \mu=0,\ldots, 5$ and
\begin{align}
P_i^{\mu} = \frac{-1}{4} \langle i^a \Sigma^{\mu} i_a \rangle = \frac{-1}{4} [ i_{\dot{a}} \tilde{\Sigma}^{\mu} i^{\dot{a}} ] \; ,
\end{align}
with \(\Sigma^{\mu}\) and \(\tilde \Sigma^{\mu}\) being the generators of
the Clifford algebra, $\Sigma^{\mu} \tilde\Sigma^{\nu} +\Sigma^{\nu} \tilde\Sigma^{\mu}=2 g^{\mu\nu}$.
Additionally one can construct spinor products \(\langle i_a j_{\dot{b}} ]\) with the property
\begin{align}
\det \big( \langle i_a j_{\dot{b}} ] \big) &= ( P_i + P_j )^2,
\end{align}
where the determinant is taken over the little group indices.

Also mirroring the four-dimensional case, we can construct a set of polarization vectors valid for massless six-momenta
\begin{align}
\varepsilon_{a \dot{a}}^{\mu}(P,K) = \frac{-1}{\sqrt{2}} \frac{\langle P_a \Sigma^{\mu} K^b \rangle \langle K_b P_{\dot{a}} ] }{2 P \! \cdot \! K} = \frac{1}{\sqrt{2}} \frac{\langle P_a K_{\dot{b}} ] [ K^{\dot{b}} \tilde{\Sigma}^{\mu} P_{\dot{a}} ] }{2 P \! \cdot \! K} ,
\label{polvector}
\end{align}
where \(K\) is an arbitrarily chosen reference vector.
The polarization vectors obey \(P \cdot \varepsilon_{a \dot{a}} = K \cdot \varepsilon_{a \dot{a}} = 0\) and the relations
\begin{align}
\varepsilon_{11} \cdot \varepsilon_{22} &= -1 \,, & \varepsilon_{12} \cdot \varepsilon_{21} &= 1 \,, & \tekst{other combinations} &= 0 \,.
\end{align}
The six-dimensional helicity sum reads,
\begin{align}
\varepsilon^{\mu}_{a\dot{b}} {\varepsilon^{\nu}}^{a\dot{b}}=\varepsilon^{\mu}_{11}\varepsilon^{\nu}_{22} + \varepsilon^{\mu}_{22}\varepsilon^{\nu}_{11} - \varepsilon^{\mu}_{12}\varepsilon^{\nu}_{21} - \varepsilon^{\mu}_{21}\varepsilon^{\nu}_{12} &= -g^{\mu \nu} + \frac{P^{\mu} K^{\nu} + K^{\mu} P^{\nu}}{P \cdot K} \; .
\label{completion}
\end{align}

The four possible combinations of little-group indices on the polarization vectors in
eq. (\ref{polvector}) correspond to the four polarization directions, or helicities, available in six
dimensions.

\subsection{Generalized unitarity in $D$ dimensions \label{sec:unitarity}}

\begin{figure}[h]
  \begin{center}
    \includegraphics[width=12cm]{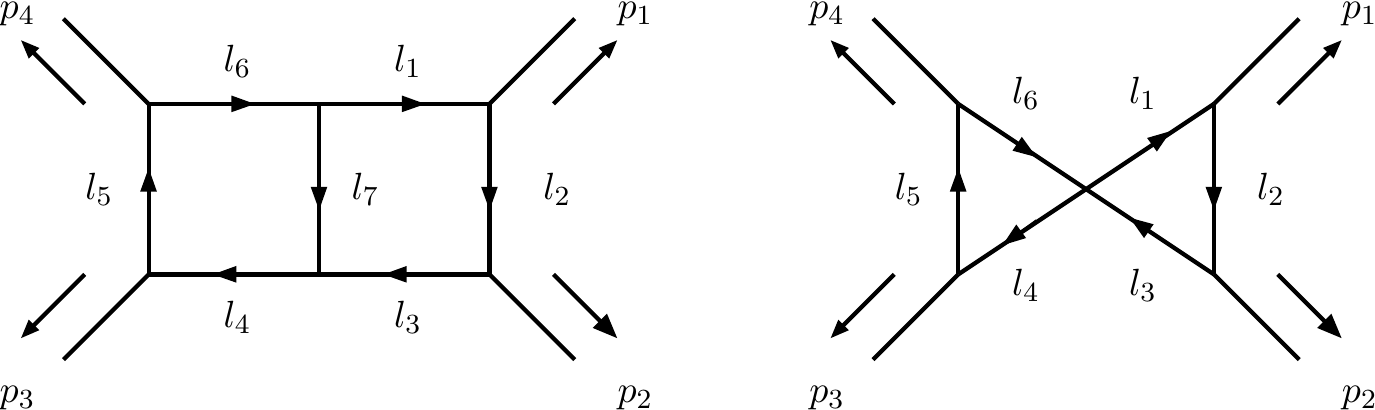}
  \end{center}
  \caption{Conventions for the momentum flow in the four-point double box, $(331)$, and the butterfly
  topology, $(330)$.}
  \label{fig:4ptParent}
\end{figure}

In this section we will illustrate our method by reproducing a known result, the two-loop
correction to the all-plus helicity amplitude for \(2 \rightarrow 2\) gluon scattering, first calculated in
\cite{Bern:2000dn}. The parent topology for this amplitude is the double-box, which can be written as
\begin{align}
  I^{[D]}_{331} = \int \frac{\id^D k_1}{(2 \pi)^D} \int \frac{\id^D k_2}{(2 \pi)^D}
  \frac{\Delta^{[D]}_{331}}{\prod\limits_{i=1}^7 l_i^2},
\end{align}
where, as shown in figure \ref{fig:4ptParent}, the propagators are,
\begin{align}
l_1 &= k_1\,, & l_2 &= k_1-p_1\,, & l_3 &= k_1-p_2-p_3\,, & l_4 &= p_3+p_4-k_2\,, \nn \\
l_5 &= p_4-k_2\,, & l_6 &= -k_2 \,,& l_7 &= -k_1-k_2\,.
\end{align}
As the topology has three propagators parametrized only by
\(k_1\), three propagators parametrized only by \(k_2\), and one parametrized in terms of both, we
denote this topology \((331)\). The only other topology contributing to the amplitude is a four leg
butterfly, which we denote \((330)\).

Using constraints from the division over the Gr\"obner basis described in section \ref{sec:integrandreduction}, as automated by the Mathematica package
{\sc BasisDet} \cite{Zhang:2012ce}, we get that the irreducible numerator \(\Delta_{331}\) can be
expressed as a sum of at most \(160\) coefficients multiplying various powers of the ISPs
\begin{align}
(k_1 \! \cdot \! \omega) \, , \;\; (k_2 \! \cdot \! \omega) \, , \;\; (k_1 \! \cdot \! p_4) \, , \;\; (k_2 \! \cdot \! p_1) \, , \;\; \mu_{11} \, , \;\; \mu_{12} \, , \;\; \mu_{22} \,,
\end{align}
where
\begin{align}
\omega^{\mu} &= \frac{\A23 \B31}{s_{12}} \frac{\cv12}{2} - \frac{\A13 \B32}{s_{12}} \frac{\cv21}{2},
\label{omegadef222}
\end{align}
is an auxiliary vector perpendicular to all the external momenta. (Since
all four external momenta are four-dimensional and linearly dependent,
there exists such an $\omega$ in four dimensions which can be explicitly
constructed using the usual four-dimensional spinor helicity formalism.) 
That is
\begin{align}
\Delta_{331} &= \sum_{i=1}^{160} c_i (k_1 \! \cdot \! \omega)^{n_{1,i}} (k_2 \! \cdot \! \omega)^{n_{2,i}} (k_1 \! \cdot \! p_4)^{n_{3,i}} (k_2 \! \cdot \! p_1)^{n_{4,i}} \mu_{11}^{n_{5,i}} \mu_{12}^{n_{6,i}} \mu_{22}^{n_{7,i}}.
\label{fourpointdelta}
\end{align}
Such a parametrization can be found for each of the three terms \(\Delta^{[6]}_g\),
\(\Delta^{[6]}_s\), and \(\Delta^{[6]}_{2s}\), contributing
to the regulated amplitude in eq. (\ref{scalareqn}), and the
specific expressions can be found using generalized unitarity cuts.

In this case we have seven on-shell constraints,
\begin{align}
  \{l_i^2 = 0\},
\label{fourpointconstraints}
\end{align}
which define a system with four degrees of freedom. As proven in section \ref{sec:integrandreduction},
this system has exactly one solution, which we choose to parametrize in terms of four free variables \(\tau_1,\ldots,\tau_4\),
\begin{align}
\bar{k}_1^{\mu} &= p_1^{\mu} + \tau_1 \frac{\A23}{\A13} \frac{\cv12}{2} + \tau_2 \frac{\B23}{\B13} \frac{\cv21}{2}, \nn \\
\bar{k}_2^{\mu} &= p_4^{\mu} + \tau_3 \frac{\A41}{\A31} \frac{\cv34}{2} + \tau_4 \frac{\B41}{\B31} \frac{\cv43}{2}.
\label{fourpointpara}
\end{align}
The three extra-dimensional parameters are determined by substituting the above expressions into,
\begin{align}
\mu_{11} &= \bar{k}_1^2 \,, & \mu_{22} &= \bar{k}_2^2 \,, & \mu_{12} &= 2 \bar{k}_1 \! \cdot \! \bar{k}_2 \,.
\label{murests}
\end{align}
Inserting eq. (\ref{fourpointpara}) into eq. (\ref{fourpointdelta}), we get an expansion
\begin{align}
\Delta_{331}|_{\tekst{cut}} &= \sum_{j=1}^{160} d_j \tau_1^{n_{1,j}} \tau_2^{n_{2,j}} \tau_3^{n_{3,j}} \tau_4^{n_{4,j}}.
\label{fourpointd}
\end{align}
for each of \(\Delta^{[6]}_g\), \(\Delta^{[6]}_s\), and \(\Delta^{[6]}_{2s}\), where the coefficients are related by
\begin{align}
{\bold d} &= M {\bold c}
\label{disstillmc}
\end{align}
where the matrix \(M\) is square and non-singular. \(\Delta_{331}|_{\tekst{cut}}\) can be found as a product
of trees, as in the four-dimensional case \cite{Britto:2004nc, Forde:2007mi, Ossola:2006us,
Badger:2012dp}. Whenever a cut gluon propagator appears, we need to sum over all possible values of
its helicity: \(\{11\}\), \(\{12\}\), \(\{21\}\), and \(\{22\}\). However, in order to correctly reproduce
the spin sum of eq. \eqref{completion}, we need to multiply by \(-1\) whenever the propagating
gluon has helicity \(\{12\}\) or \(\{21\}\).

\begin{figure}[!tp]
       \centering
       \begin{subfigure}{4.5cm}
               \centering
               \includegraphics[width=4.5cm]{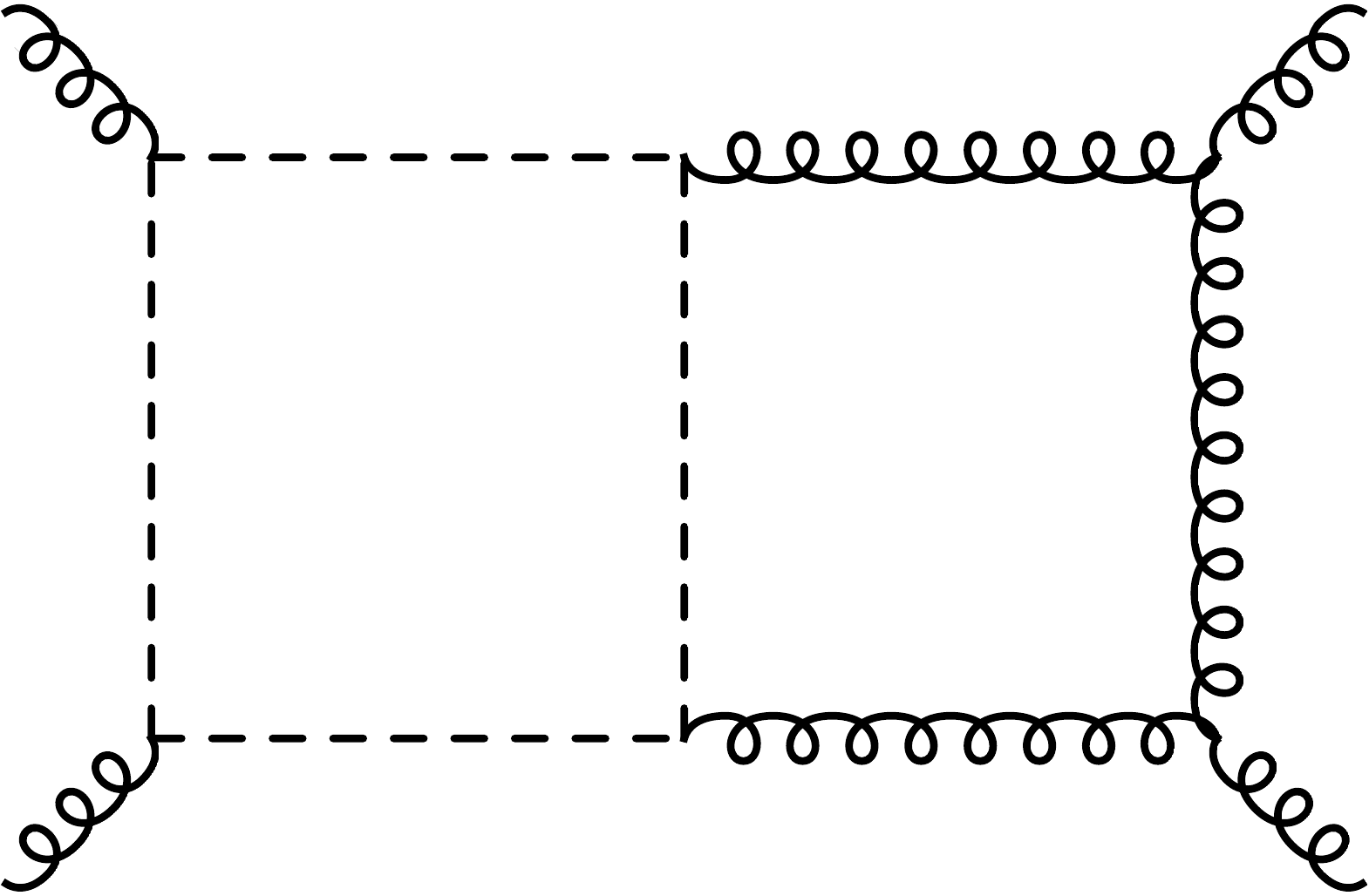}
       \end{subfigure}\quad
       \begin{subfigure}{4.5cm}
               \centering
               \includegraphics[width=4.5cm]{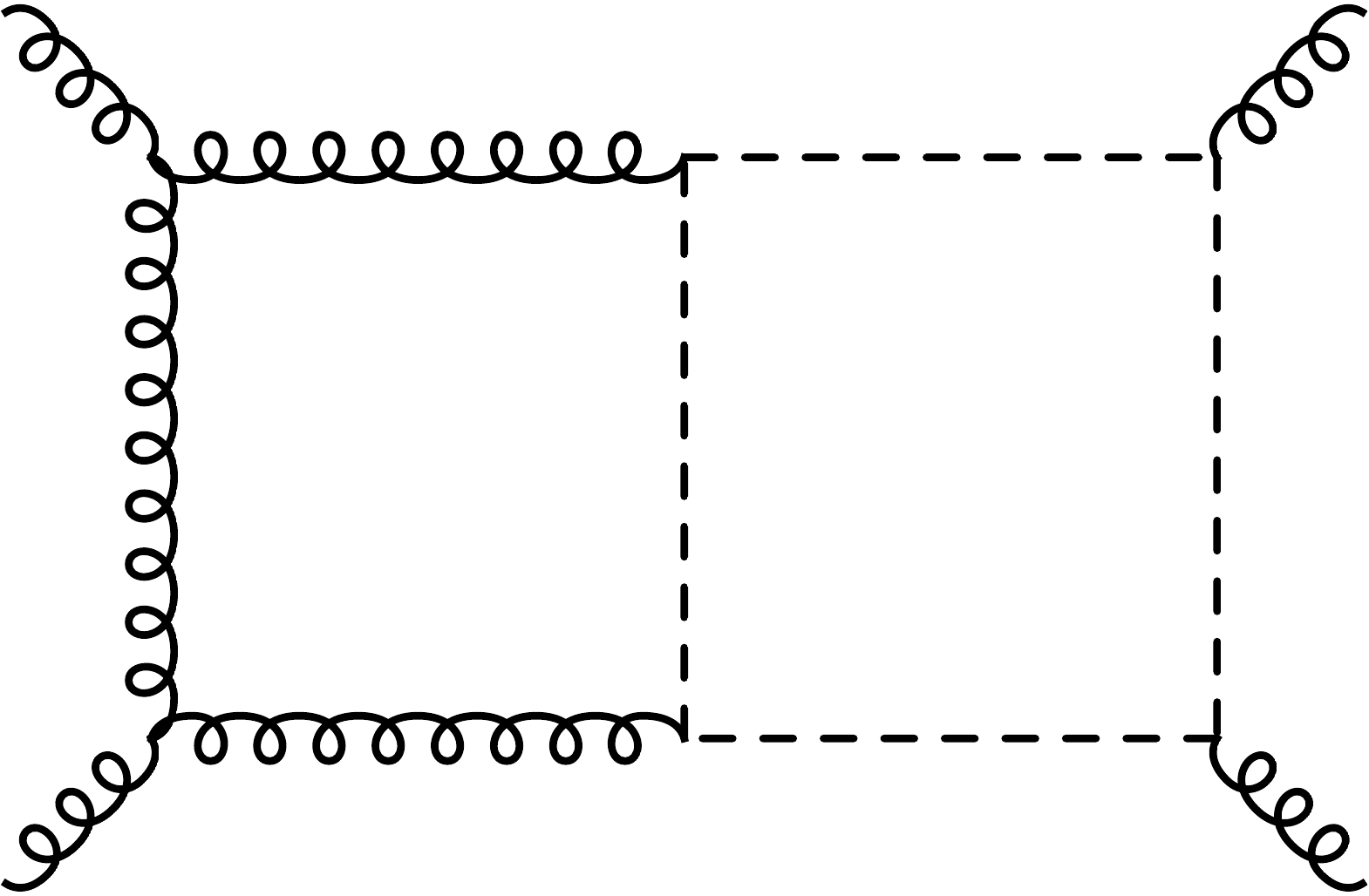}
       \end{subfigure}\quad
       \begin{subfigure}{4.5cm}
               \centering
               \includegraphics[width=4.5cm]{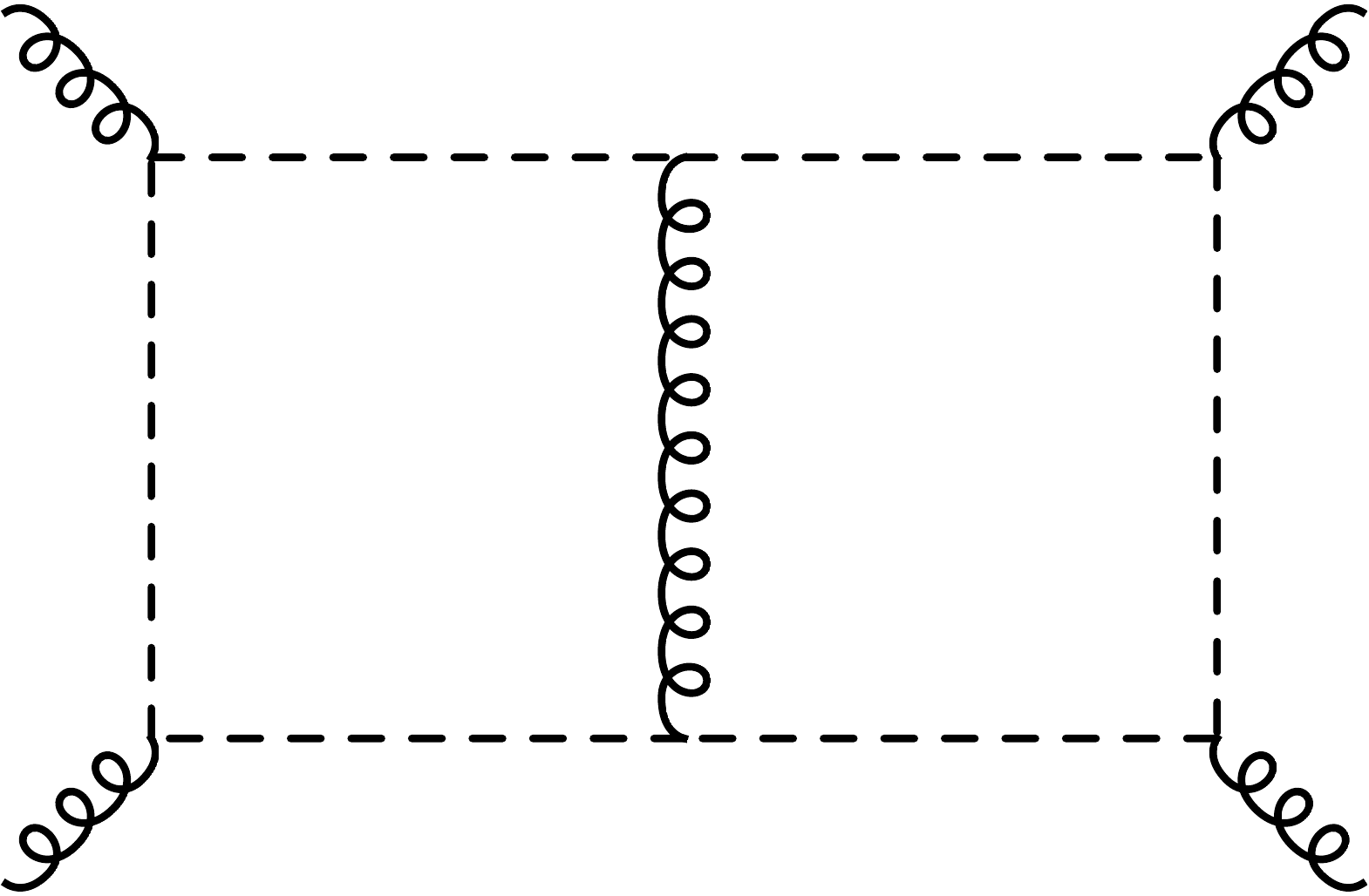}
       \end{subfigure}
       \caption{The flavour contributions to \(\Delta_{s,\;331}^{[6]}\).}
       \label{fig:flavour331}
\end{figure}

For the \(\Delta^{[6]}_g\)-contribution we get for instance
\begin{align}
\Delta_{g,\;331}^{[6]}&\big|_{\tekst{cut}} = \nonumber\\&
\sum_{ \substack{ \{h_1,\ldots,h_7\} \in \\ \{11,12,21,22\} }} \!\!\!\! \bigg( \sigma_{h_1,\ldots,h_7}
\kur{A}(-l_1^{-h_1}, p_1^{(11)}, l_2^{h_2} )
\kur{A}(-l_2^{-h_2}, p_2^{(11)}, l_3^{h_3} )
\kur{A}(-l_3^{-h_3}, l_4^{h_4}, -l_7^{-h_7} ) \nonumber\\&\hspace{5mm}
\kur{A}(-l_4^{-h_4}, p_3^{(11)}, l_5^{h_5} )
\kur{A}(-l_5^{-h_5}, p_4^{(11)}, l_6^{h_6} )
\kur{A}(-l_6^{-h_6}, l_1^{h_1}, l_7^{h_7} )
\bigg),
\label{gluoncut}
\end{align}
with
\begin{align}
\sigma_{h_1,\ldots,h_n} &= \prod_{i=1}^n (2\delta_{a_i \dot{a}_i}-1),
\end{align}
and with the \(\{11\}\) on the external gluons corresponding to helicity `plus' in four dimensions.

\(\Delta^{[6]}_{s,\;331}\) contains three contributions, each having one complete scalar loop as shown
in fig. \ref{fig:flavour331}. Each term can be constructed similarly to eq. \eqref{gluoncut}, with no subtleties
arising from cut scalars. \(\Delta^{[6]}_{2s}\) is zero for the $(331)$ topology.
Having extracted the values of the coefficients $d_j$ in eq. \eqref{fourpointd} for both contributions, we are able
to solve the linear system in eq. \eqref{disstillmc} and compute \(\Delta_{331}\) from eq. \eqref{scalareqn}:
\begin{align}
  \Delta_{g,\;331}^{[D_s]} &= \Delta_{g,\;331}^{[6]} + (D_s-6) \Delta_{s,\;331}^{[6]}.
\end{align}
Following this procedure, one can check the known result of ref. \cite{Bern:2000dn}:
\begin{align}
  \Delta_{331} &= \frac{-is_{12}^2 s_{14} F_1 (D_s,\mu_{11},\mu_{22},\mu_{12})}{\A12 \A23 \A34 \A41},
\label{doublebox222}
\end{align}
where
\begin{align}
  F_1(D_s,\mu_{11},\mu_{22},\mu_{12})
  = (D_s-2)\left( \mu_{11}\mu_{22} + \mu_{11}\mu_{33} + \mu_{22}\mu_{33} \right) + 4\left(
  \mu_{12}^2 - 4\mu_{11}\mu_{22} \right),
\label{fdef}
\end{align}
and $\mu_{33} = \mu_{11} + \mu_{22} + \mu_{12}$.

For the butterfly topology \((330)\) all the above can be repeated. The solution to the on-shell
constraints are identical to eqs. \eqref{fourpointpara} and \eqref{murests}, with the exception that no constraint fixes
\(\mu_{12}\), making it a fifth free parameter \(\mu_{12} = s_{12} \tau_5\), where \(s_{12}\) has
been inserted to make \(\tau_5\) dimensionless.  The ISPs are the same as for the \((331)\)-case,
though in this case the most general ISP basis has \(146\) terms.

\begin{figure}[!tp]
       \centering
       \begin{subfigure}{3.2cm}
               \centering
               \includegraphics[width=3.2cm]{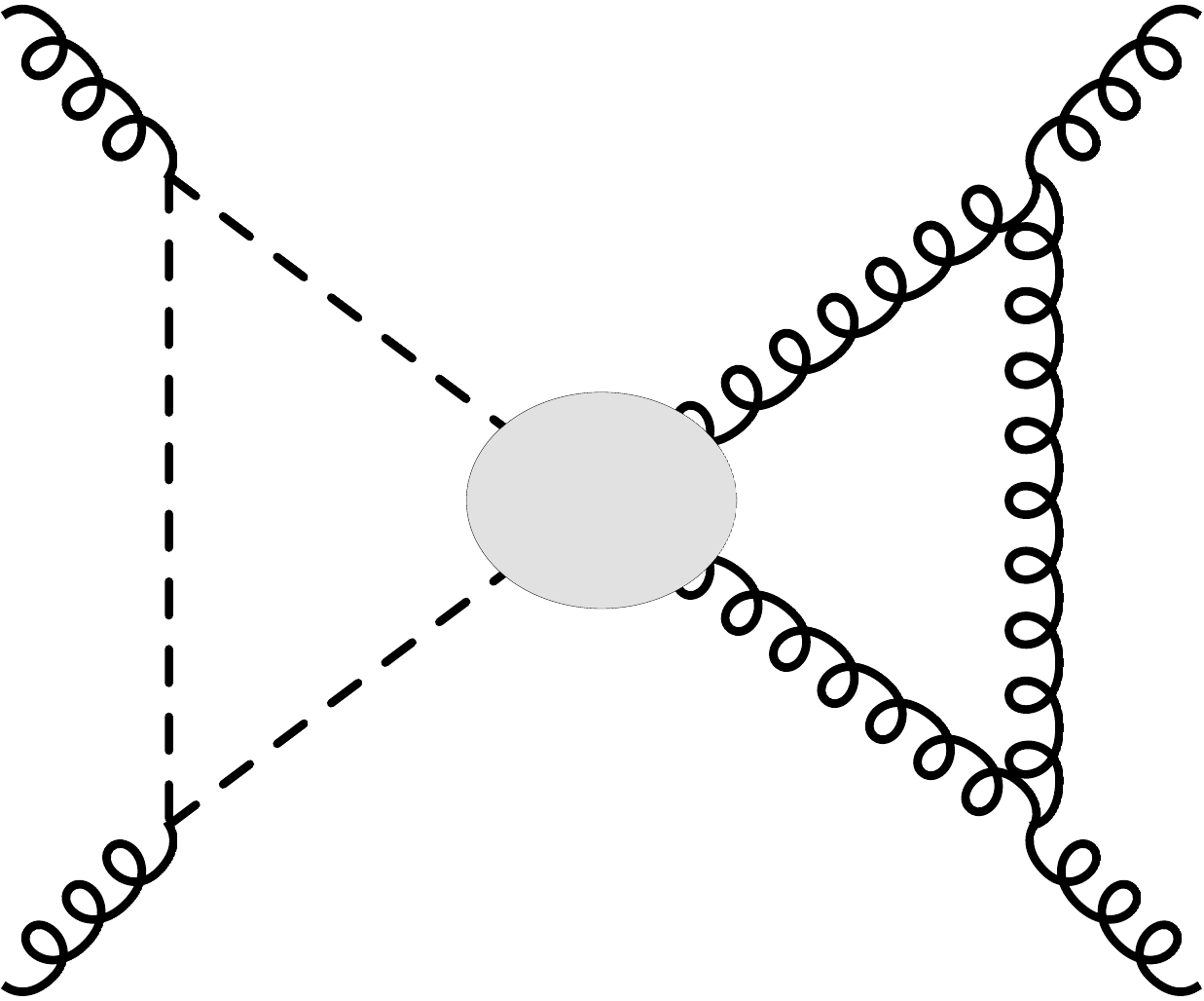}
       \end{subfigure}\quad
       \begin{subfigure}{3.2cm}
               \centering
               \includegraphics[width=3.2cm]{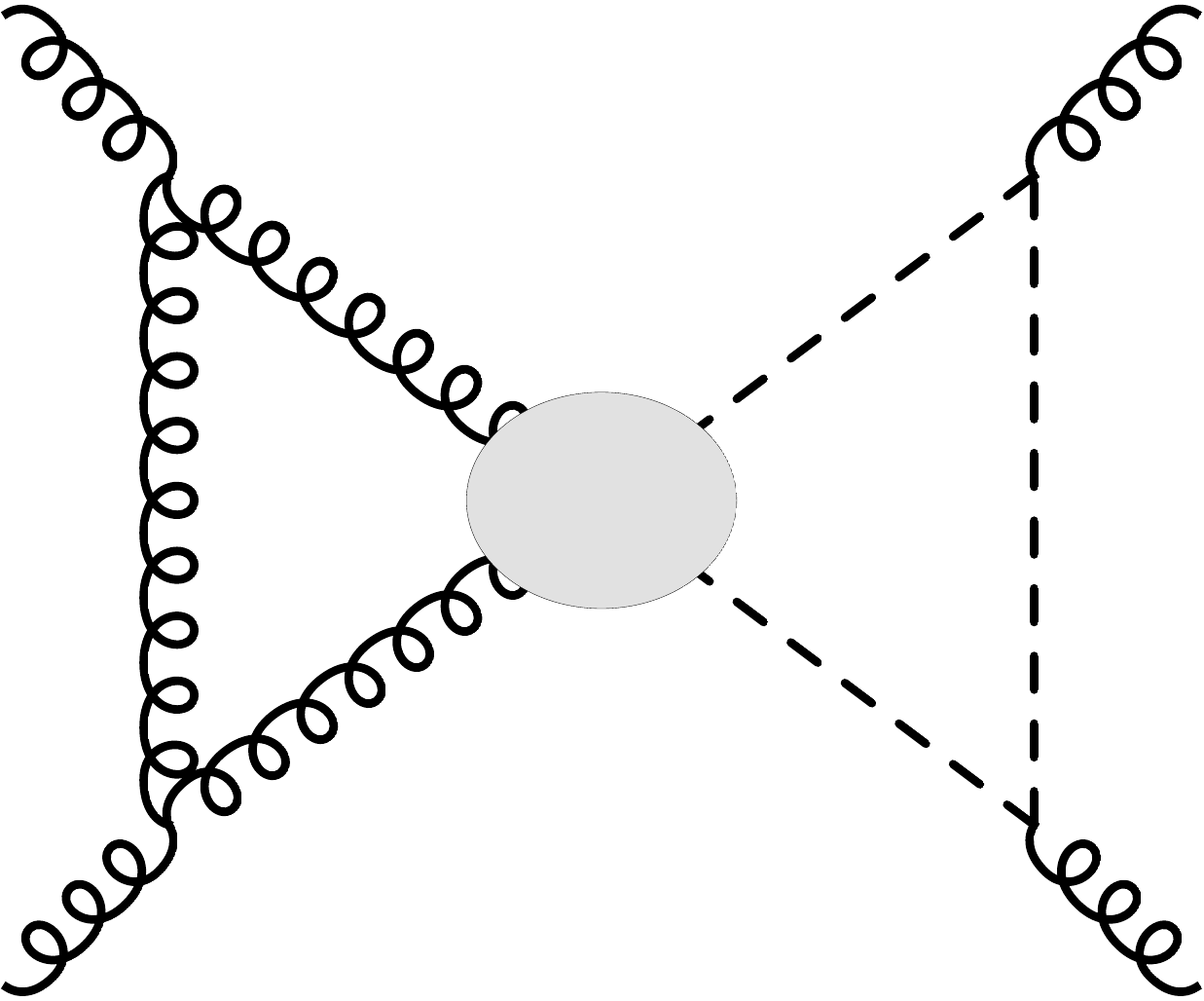}
       \end{subfigure}\quad
       \begin{subfigure}{3.2cm}
               \centering
               \includegraphics[width=3.2cm]{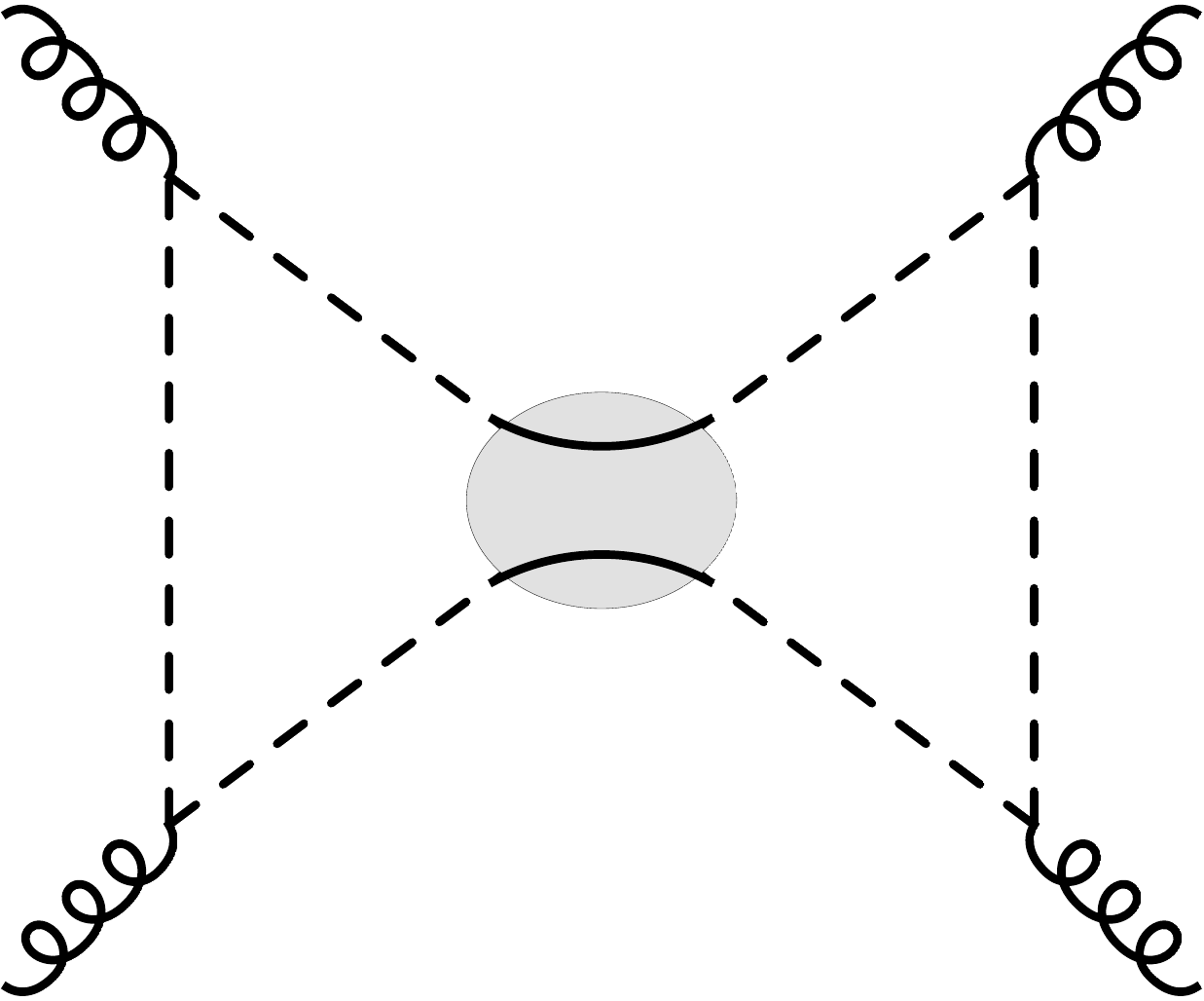}
       \end{subfigure}\quad
       \begin{subfigure}{3.2cm}
               \centering
               \includegraphics[width=3.2cm]{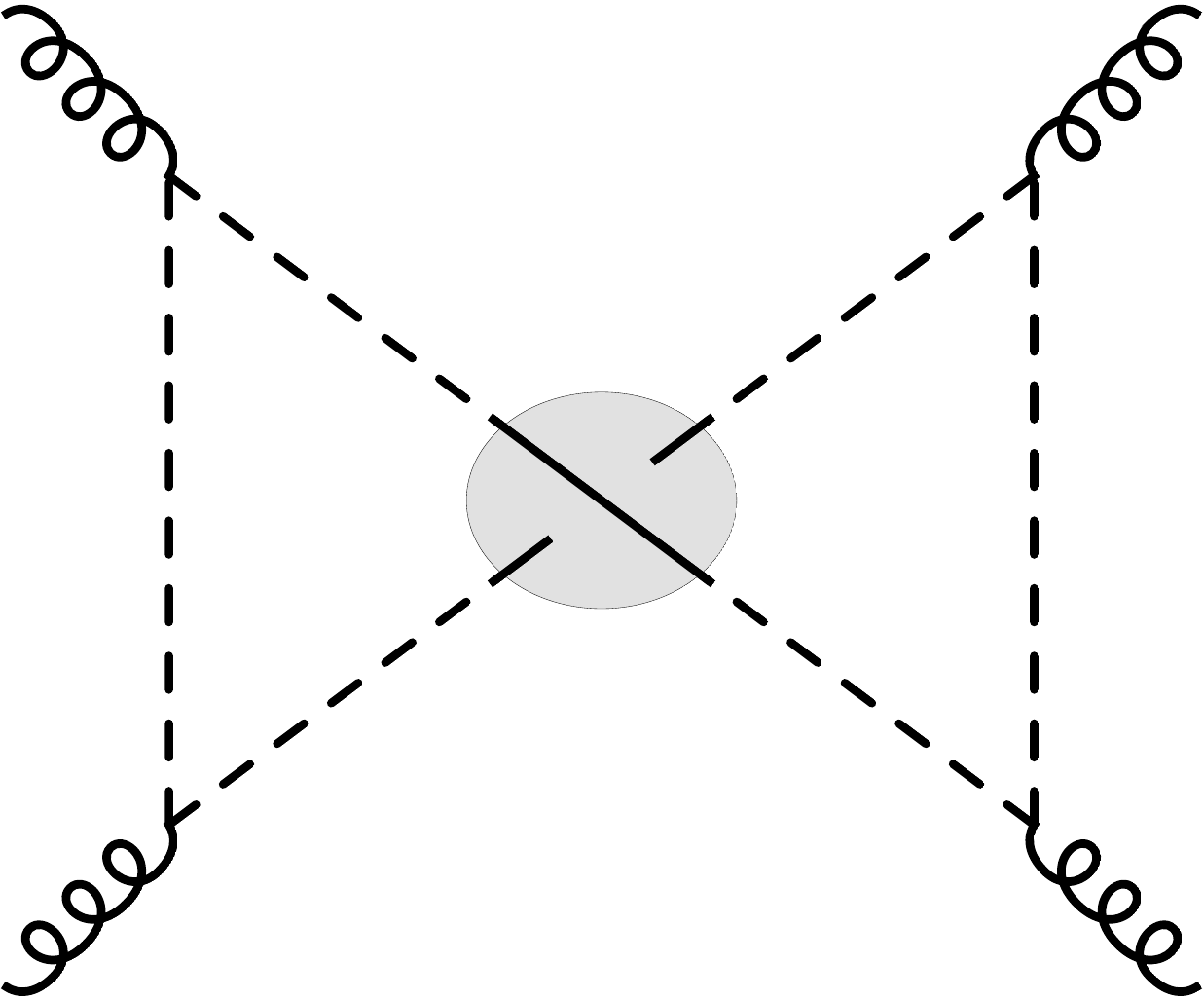}
       \end{subfigure}
       \caption{The flavour contributions to \(\Delta_{s,\;330}^{[6]}\).}
       \label{fig:flavour330}
\end{figure}

\begin{figure}[!bp]
       \centering
       \begin{subfigure}{3.2cm}
               \centering
               \includegraphics[width=3.2cm]{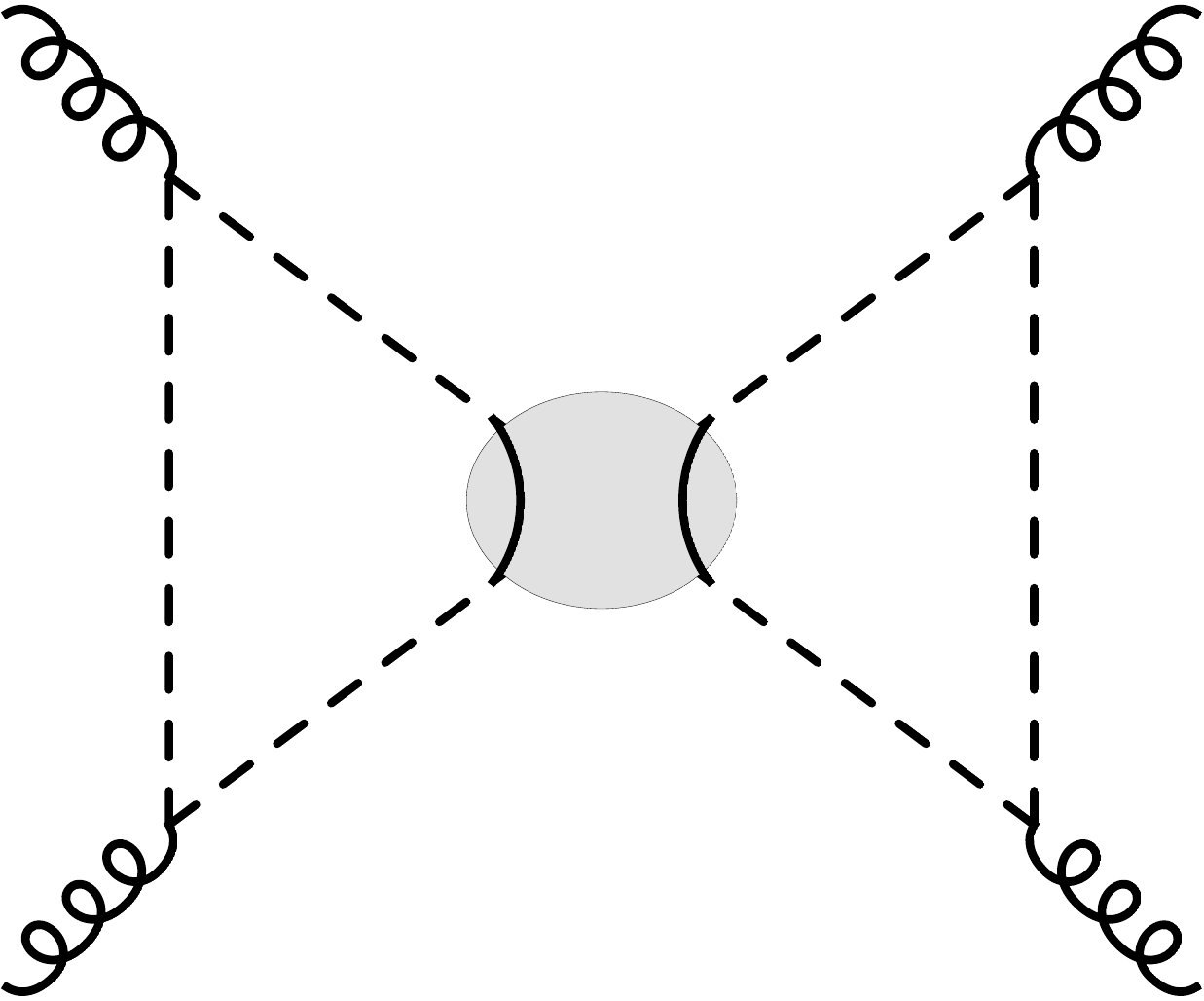}
       \end{subfigure}\quad
       \caption{The flavour contribution to \(\Delta_{2s,\;330}^{[6]}\).}
       \label{fig:flavours2}
\end{figure}

When performing this sub-maximal cut we must remove the previously computed leading singularity,
$(331)$, using the OPP subtraction procedure. For \(\Delta_g^{[6]}\) the cut integrand is simply,
\begin{align}
\Delta_{g,\;330}^{[6]}&\big|_{\tekst{cut}} = \nonumber\\&
\!\!\! \sum_{ \{h_1,\ldots,h_6\} } \!\!\! \bigg( \sigma_{h_1,\ldots,h_6}
  \kur{A}(-l_1^{-h_1}, p_1^{(11)}, l_2^{h_2} )
  \kur{A} (-l_2^{-h_2}, p_2^{(11)}, l_3^{h_3} )
  \kur{A} (-l_3^{-h_3}, l_4^{h_4}, -l_6^{-h_6}, l_1^{h_1} ) \nonumber\\&\hspace{5mm}
  \kur{A}(-l_4^{-h_4}, p_3^{(11)}, l_5^{h_5} )
  \kur{A}(-l_5^{-h_5}, p_4^{(11)}, l_6^{h_6} )
\bigg) - \frac{i}{(l_6-l_1)^2} \Delta_{g,\;331}^{[6]} .
\end{align}

\(\Delta_{s,\;330}^{[6]}\), shown in figure \ref{fig:flavour330}, will have the same three contributions as the \((331)\)-topology, but in
addition it will get a fourth contribution coming from the four-point scalar \(ss'ss'\)-vertex (see
appendix \ref{feynmanappendix}), which can be interpreted as a scalar loop forming a figure eight.
\(\Delta_{2s,\;330}^{[6]}\), shown in figure \ref{fig:flavours2}, is non-zero and comes from the
\(sss's'\)-vertex.


After solving the linear system to find the integrand coefficients, the full result, via eq.
\eqref{scalareqn}, can be shown to reproduce the known result \cite{Bern:2000dn},
\begin{align}
\Delta_{330} &= \frac{-is_{12} s_{14} \Big( 2 (D_s-2) (\mu_{11} + \mu_{22}) \mu_{12} + (D_s-2)^2 \mu_{11} \mu_{22} \big( (k_1+k_2)^2 + s_{12} \big)/s_{12} \Big)}{\A12 \A23 \A34 \A41}.
\end{align}

\subsection{Feynman diagram set-up for the FDH scheme \label{sec:feynman}}

The input for the reduction can be generated from Feynman diagrams
using a Feynman gauge for the internal propagators and using,
\begin{align}
  g^{\mu}_{\mu} = D_s,
  \label{eq:FDspinsum}
\end{align}
where $D_s>D=4-2\eps$. To correctly reproduce the results of the four-dimensional helicity scheme
obtained from the six-dimensional formalism via eq. \eqref{scalareqn},
a four ghost interaction is necessary in butterfly-type topologies \cite{Bern:2002zk}. We find the momentum
twistor parametrization outlined in appendix \ref{app:momtwistor} particularly useful when
dealing with the large intermediate expressions that arise from this approach.

\section{Integrand reduction for the five gluon amplitude}

In this section we summarize the generalized unitarity cuts for all topologies contributing to the
all-plus helicity two-loop gluon amplitude in Yang-Mills theory.
\begin{figure}[!tp]
  \centering
       \begin{subfigure}{3.2cm}
               \centering
               \includegraphics[width=3.2cm]{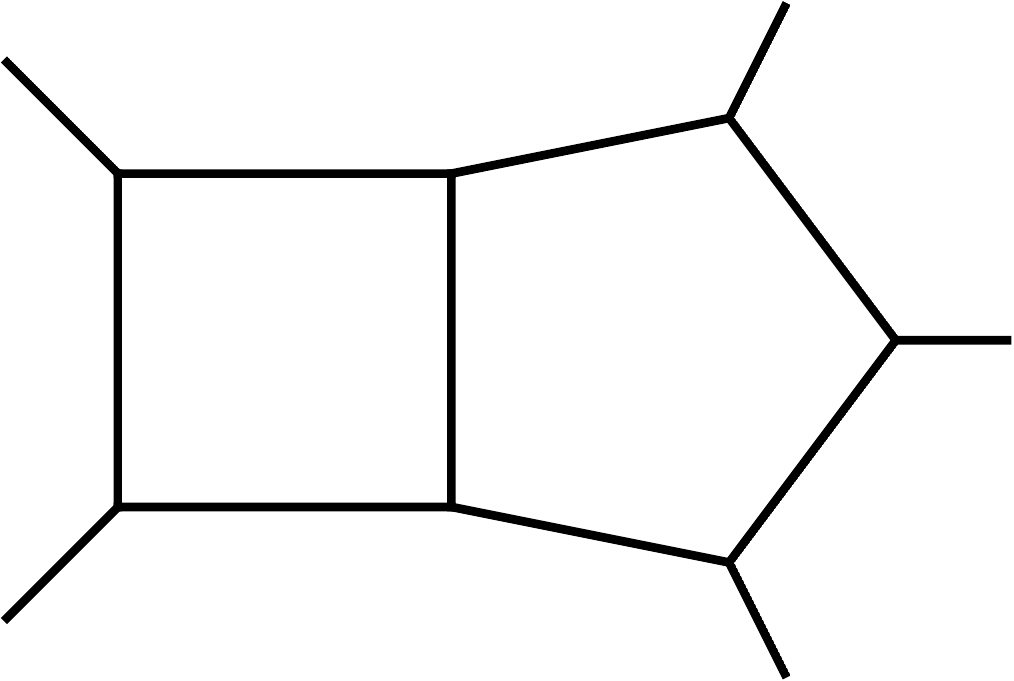}
       \end{subfigure}\quad
       \begin{subfigure}{3.2cm}
               \centering
               \includegraphics[width=3.2cm]{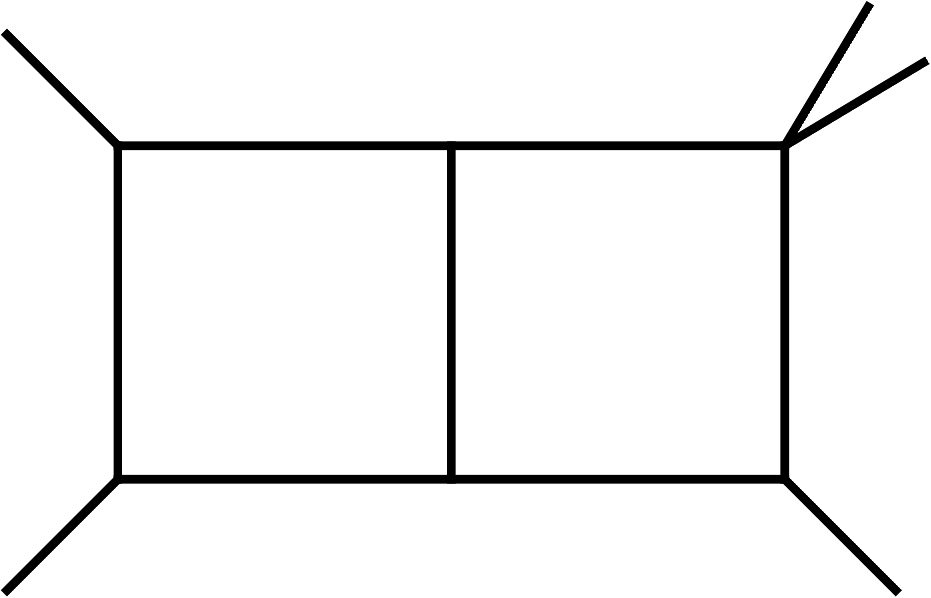}
       \end{subfigure}\quad
       \begin{subfigure}{3.2cm}
               \centering
               \includegraphics[width=3.2cm]{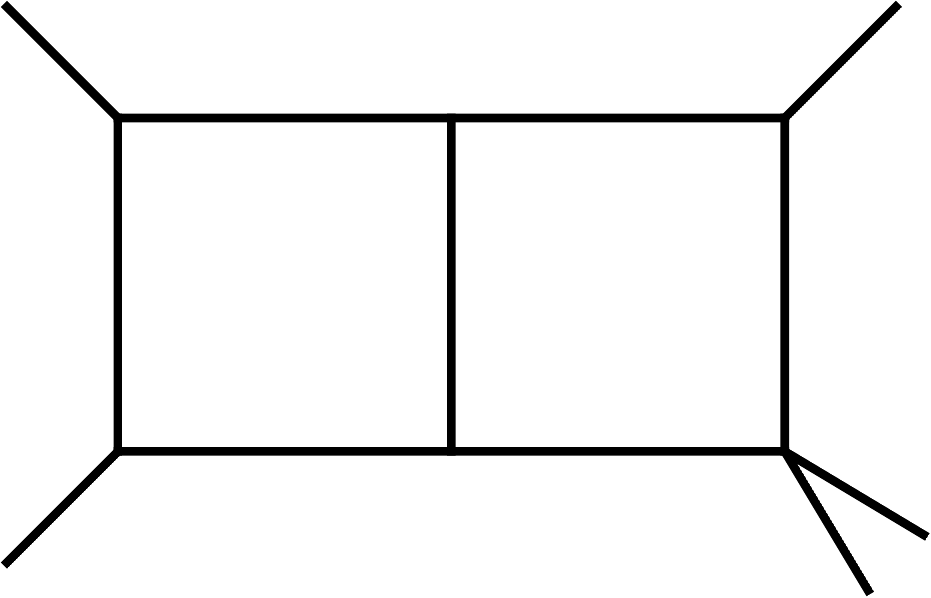}
       \end{subfigure}\quad
       \begin{subfigure}{3.2cm}
               \centering
               \includegraphics[width=3.2cm]{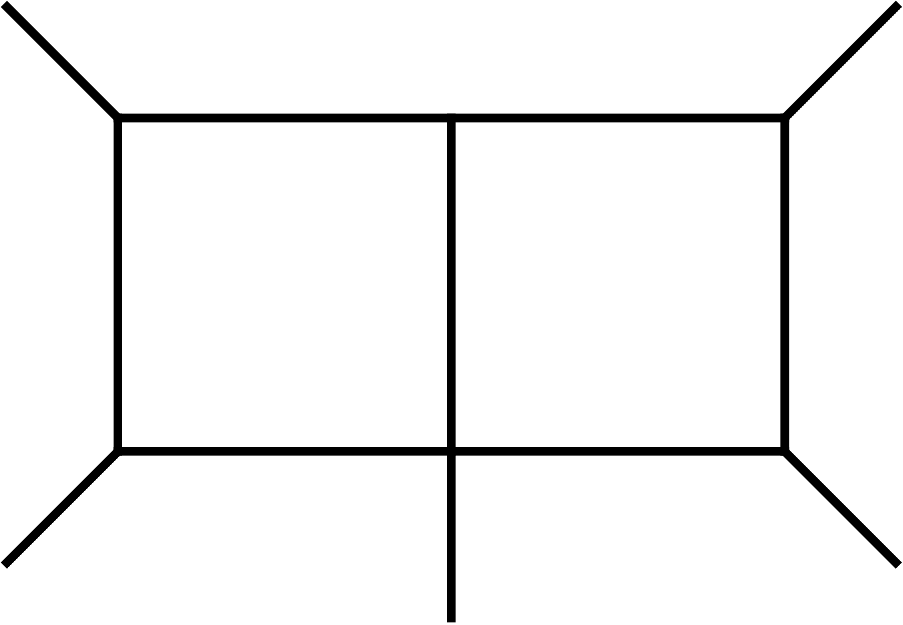}
       \end{subfigure}\quad
       \begin{subfigure}{3.2cm}
               \centering
               \includegraphics[width=3.2cm]{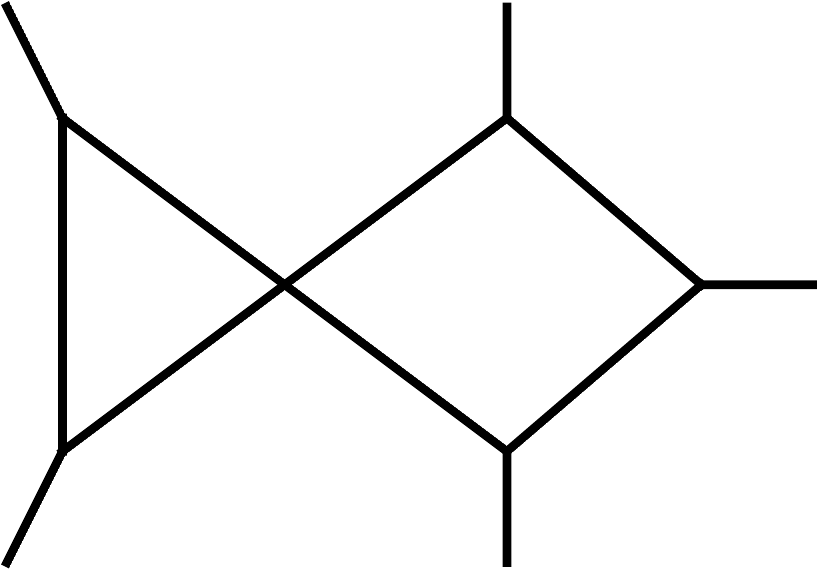}
       \end{subfigure}\quad
       \begin{subfigure}{3.2cm}
               \centering
               \includegraphics[width=3.2cm]{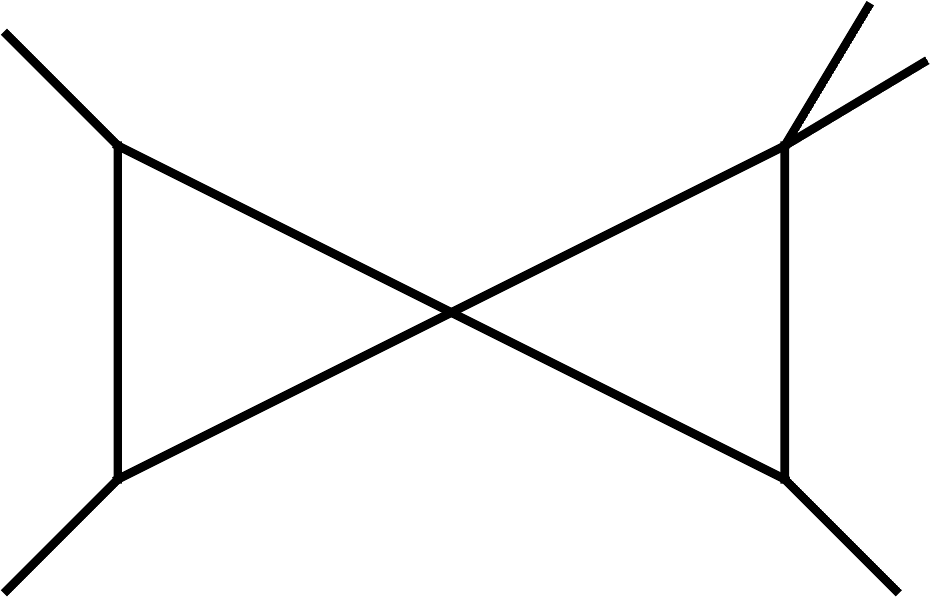}
       \end{subfigure}\quad
       \begin{subfigure}{3.2cm}
               \centering
               \includegraphics[width=3.2cm]{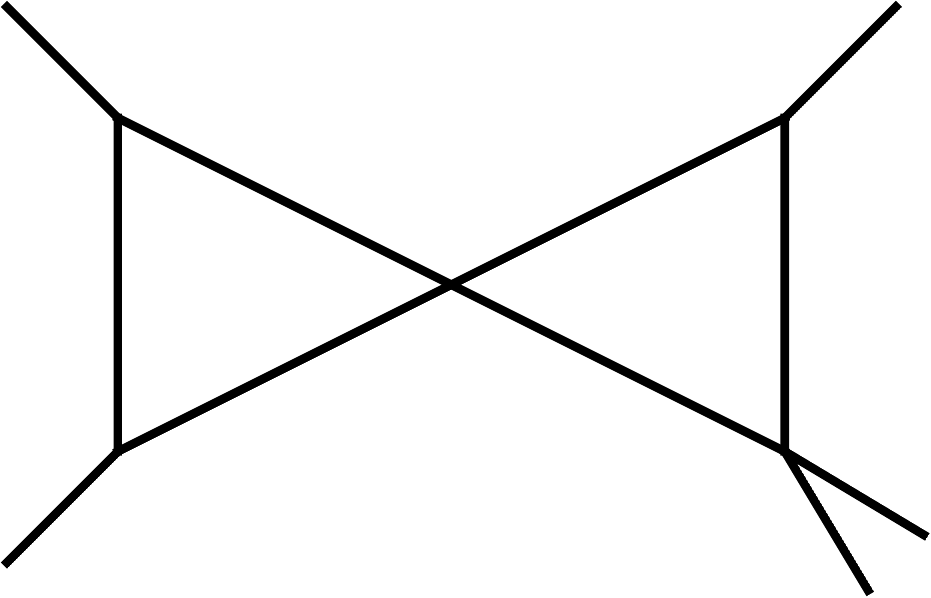}
       \end{subfigure}\quad
       \begin{subfigure}{3.2cm}
               \centering
               \includegraphics[width=3.2cm]{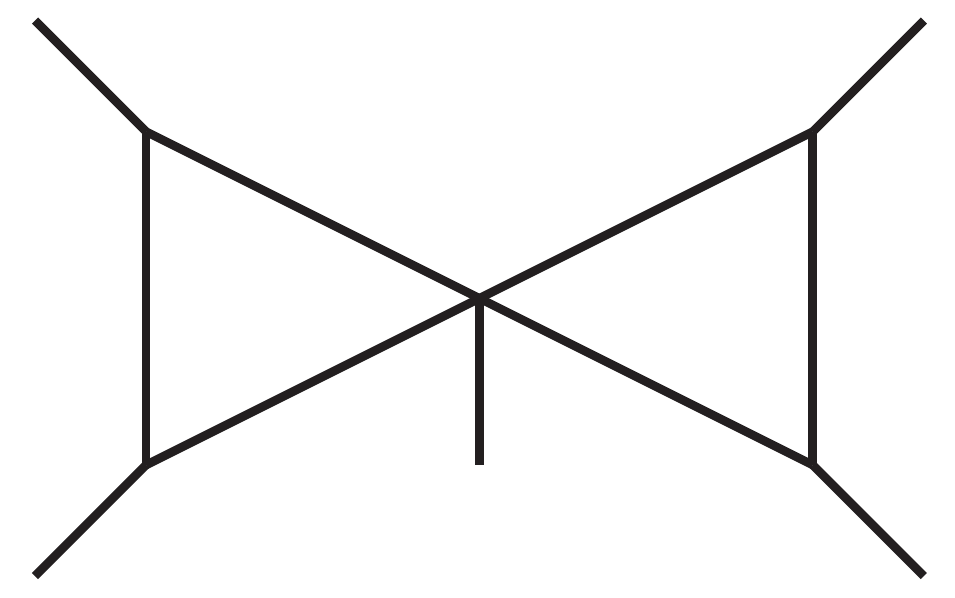}
       \end{subfigure}\quad
       \caption{The eight topologies. From upper left to lower right: (431), (331; M\(_1\)), (331; M\(_2\)), (331; 5L), (430), (330; M\(_1\)), (330; M\(_2\)), (330; 5L).}
       \label{figtopologies}
\end{figure}
It turns out that only eight different topologies, each with six or more propagators,
are necessary to write down the complete amplitude (fig. \ref{figtopologies}).
In addition to those described in detail below, we have computed a selection of four, five and
six propagator cuts which all evaluate to zero. These cuts have been important in determining the exact
form of the tensor integrals presented in this section.
\begin{figure}[!bp]
  \centering
  \includegraphics[width=7cm]{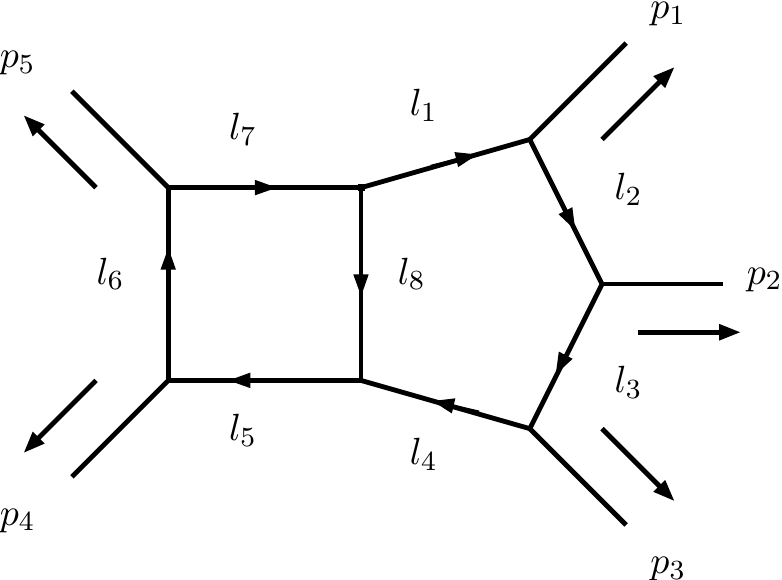}
  \caption{Conventions for the momentum flow in the pentagon-box parent topology, $(431)$.}
  \label{fig:5ptParent}
\end{figure}
All the eight topologies are descended from the same maximum cut parent topology $(431)$, (fig. \ref{fig:5ptParent}) which has eight propagators:
\begin{align}
  l_1 &= k_1, & l_2 &= k_1-p_1, & l_3 &= k_1-p_1-p_2, \nn \\
  l_4 &= k_1-p_1-p_2-p_3, & l_5 &= -k_2+p_4+p_5, & l_6 &= -k_2+p_5, \nn \\
  l_7 &= -k_2, & l_8 &= -k_1-k_2,
  \label{eq:2to3props}
\end{align}
meaning that the propagators of each topology are a subset of the eight above.

Only six of the topologies are independent, as the $M_1$ and $M_2$ topologies are related by
\begin{align}
  \Delta_{M_1}(k_1,k_2,p_1,p_2,p_3,p_4,p_5) &= -\Delta_{M_2}(-p_{45}-k_1,p_{45}-k_2,p_3,p_2,p_1,p_5,p_4),
\end{align}
and in the following sections we will present the on-shell cut solutions and integrand
parameterizations for these six topologies, along with a list of ISPs and the allowed number of ISP monomials.

\subsection{The pentagon-box: $(431)$}

The maximum eight-fold cut can be parametrized using three free parameters $\tau_1, \tau_2$ and $\tau_3$, as
\begin{align}
  \bar{k}_1^\mu &= a_1 p_1^\mu + a_2 p_2^\mu + a_3 \tfrac{\A23}{2\A13}\cv12 + a_4 \tfrac{\B23}{2\B13}\cv21, \nn \\
  \bar{k}_2^\mu &= b_1 p_4^\mu + b_2 p_5^\mu + b_3 \tfrac{\A51}{2\A41}\cv45 + b_4 \tfrac{\B51}{2\B41}\cv54,
  \label{eq:oseqs-db320}
\end{align}
where
\begin{align}
  a_1 &= 1 \,, &
  a_2 &= 0 \,, &
  a_3 &= \tau_1 \,, &
  a_4 &= 1-\tau_1 \,, \nn \\
  b_1 &= 0 \,, &
  b_2 &= 1 \,, &
  b_3 &= \tau_2 \,, &
  b_4 &= \tau_3 \,, \nn \\
  \mu_{11} &= \bar{k}_1^2 \,, &
  \mu_{22} &= \bar{k}_2^2 \,, &
  \mu_{12} &= 2\MP{\bar{k}_1}{\bar{k}_2} \,.
\end{align}

The general integrand has $79$ coefficients in terms of the ISPs
\begin{align}
(k_1 \cdot p_5) \, , \,\; (k_2 \cdot p_2) \, , \,\; (k_2 \cdot p_1) \, , \,\; \mu_{11} \, , \,\; \mu_{12} \, , \,\; \mu_{22} \, ,
\end{align}
the form of which can easily be obtained using the {\sc BasisDet}
package \cite{Zhang:2012ce}. We choose to prefer the monomials in $\mu_{ij}$ over the higher
powers of $(k_i \cdot p_j)$ which would be preferred by the polynomial division. This is important in order
to make the four dimensional limit manifest.

For the all-plus helicity configuration of the five gluon amplitude we find:
\begin{align}
  &\Delta_{431}(1^+,2^+,3^+,4^+,5^+) =
  \nonumber\\&
  -\frac{i \, s_{12}s_{23}s_{45} \, F_1(D_s,\mu_{11},\mu_{22},\mu_{12})}{\A12\A23\A34\A45\A51
  \trfive}
  \left( \tr_+(1345) (k_1 + p_5)^2 + s_{15}s_{34}s_{45}
  \right),
  \label{delta431}
\end{align}
where
\begin{align}
  F_1(D_s,\mu_{11},\mu_{22},\mu_{12})
  = (D_s-2)\left( \mu_{11}\mu_{22} + \mu_{11}\mu_{33} + \mu_{22}\mu_{33} \right) + 4\left(
  \mu_{12}^2 - 4\mu_{11}\mu_{22} \right),
  \label{eq:F1def}
\end{align}
with $\mu_{33} = \mu_{11} + \mu_{22} + \mu_{12}$, just like for the
\(2 \rightarrow 2\) case in eq. (\ref{fdef}).

\subsection{The massive double-box: $(331;M_2)$}

This topology can be parametrized by
\begin{align}
  \bar{k}_1^\mu &= a_1 p_1^\mu + a_2 \flm{p_{23}} + a_3 \tfrac{\A{\fl{p_{23}}}4}{2\A14}\cv1{\fl{p_{23}}} + a_4 \tfrac{\B{\fl{p_{23}}}4}{2\B14} \cv{\fl{p_{23}}}1 , \nn \\
  \bar{k}_2^\mu &= b_1 p_4^\mu + b_2 p_5^\mu + b_3 \tfrac{\A51}{2\A41}\cv45 + b_4 \tfrac{\B51}{2\B41}\cv54 ,
  \label{eq:oseqs-db220c}
\end{align}
where
\begin{align}
  a_1 &= 1 \,, &
  a_2 &= 0 \,, &
  a_3 &= \tau_1 \,, &
  a_4 &= \tau_2 \,, \nn \\
  b_1 &= 0 \,, &
  b_2 &= 1 \,, &
  b_3 &= \tau_3 \,, &
  b_4 &= \tau_4 \,, \nn \\
  \mu_{11} &= \bar{k}_1^2 \,, &
  \mu_{22} &= \bar{k}_2^2 \,, &
  \mu_{12} &= 2\MP{\bar{k}_1}{\bar{k}_2} \,,
\end{align}
and
\begin{equation}
  \flm{p_{23}} = p_2^\mu+p_3^\mu + \frac{s_{23}}{s_{12} + s_{13}} p_1^\mu .
\end{equation}

The general integrand has $160$ coefficients in terms of the ISPs
\begin{align}
(k_1\cdot \omega) \, , \,\; (k_2\cdot \omega) \, , \,\; (k_1 \cdot p_5) \, , \,\; (k_2\cdot
p_1) \, , \,\; \mu_{11} \, , \,\; \mu_{12} \, , \,\; \mu_{22} \,,
\end{align}
with \(\omega^{\mu}\) being a vector which can be constructed to be perpendicular to $p_1$, $p_4$, $p_5$, and $p_{23}$, the specific form of which is not important.

We find the result
\begin{align}
  &\Delta_{331;M_2}(1^+,2^+,3^+,4^+,5^+) = \nonumber\\&
  -\frac{i \, s_{15}s_{45}^2  \, \tr_-(1234) \, F_1(D_s,\mu_{11},\mu_{22},\mu_{12})}{\A12\A23\A34\A45\A51 \trfive} .
  \label{eq:5ptallplus_db220c}
\end{align}

\subsection{The five-legged double-box: $(331;5L)$}

This topology can be parametrized by
\begin{align}
  \bar{k}_1^\mu &= a_1 p_1^\mu + a_2 p_2^\mu + a_3 \tfrac{\A23}{2\A13}\cv12 + a_4 \tfrac{\B23}{2\B13}\cv21 , \nn \\
  \bar{k}_2^\mu &= b_1 p_4^\mu + b_2 p_5^\mu + b_3 \tfrac{\A51}{2\A41}\cv45 + b_4 \tfrac{\B51}{2\B41}\cv54 ,
  \label{eq:oseqs-db2205L}
\end{align}
where
\begin{align}
  a_1 &= 1 \,, &
  a_2 &= 0 \,, &
  a_3 &= \tau_1\,,  &
  a_4 &= \tau_2 \,, \nn \\
  b_1 &= 0 \,, &
  b_2 &= 1 \,, &
  b_3 &= \tau_3 \,, &
  b_4 &= \tau_4 \,, \nn \\
  \mu_{11} &= \bar{k}_1^2 \,, &
  \mu_{22} &= \bar{k}_2^2 \,, &
  \mu_{12} &= 2\MP{\bar{k}_1}{\bar{k}_2} \,.
\end{align}

The general integrand has $160$ coefficients in terms of the ISPs
\begin{align}
(k_1 \cdot p_5) \, , \,\; (k_1 \cdot p_4) \, , \,\; (k_2 \cdot p_2) \, , \,\; (k_2 \cdot
p_1) \, , \,\; \mu_{11} \, , \,\; \mu_{12} \, , \,\; \mu_{22} \,,
\end{align}

and we find the result
\begin{align}
  &\Delta_{331;5L}(1^+,2^+,3^+,4^+,5^+) = \nonumber\\&
  \frac{i \, s_{12}s_{23}s_{34}s_{45}s_{15} \, F_1(D_s,\mu_{11},\mu_{22},\mu_{12})}{\A12\A23\A34\A45\A51 \trfive}.
  \label{eq:5ptallplus_db2205L}
\end{align}

\subsection{The box-triangle butterfly: $(430)$}

This topology can be parametrized by
\begin{align}
  \bar{k}_1^\mu &= a_1 p_1^\mu + a_2 p_2^\mu + a_3 \tfrac{\A23}{2\A13}\cv12 + a_4 \tfrac{\B23}{2\B13}\cv21 ,\nn \\
  \bar{k}_2^\mu &= b_1 p_4^\mu + b_2 p_5^\mu + b_3 \tfrac{\A51}{2\A41}\cv45 + b_4 \tfrac{\B51}{2\B41}\cv54 ,
  \label{eq:oseqs-bt32}
\end{align}
where
\begin{align}
  a_1 &= 1 \,, &
  a_2 &= 0 \,, &
  a_3 &= \tau_1 \,, &
  a_4 &= 1-\tau_1 \,, \nn \\
  b_1 &= 0 \,, &
  b_2 &= 1 \,, &
  b_3 &= \tau_2 \,, &
  b_4 &= \tau_3 \,, \nn \\
  \mu_{11} &= \bar{k}_1^2 \,, &
  \mu_{22} &= \bar{k}_2^2 \,, &
  \mu_{12} &= s_{45} \tau_4 \,.
\end{align}

The general integrand has $85$ coefficients in terms of the ISPs
\begin{align}
(k_1 \cdot \omega_{123}) \, , \,\; (k_2 \cdot \omega_{45-}) \, , \,\; (k_2 \cdot \omega_{45+}) \, , \,\; \mu_{11} \, , \,\; \mu_{12} \, , \,\; \mu_{22} \,,
\end{align}
with
\begin{align}
\omega_{123}^{\mu} &= \frac{\A23 \B31}{s_{12}} \frac{\cv12}{2} - \frac{\A13 \B32}{s_{12}} \frac{\cv21}{2},
\label{omegadef}
\end{align}
defined as in the \(2 \rightarrow 2\) case in eq. (\ref{omegadef222}), to be perpendicular to $p_1$, $p_2$, and $p_3$. \(\omega_{45-}\) and \(\omega_{45+}\) are defined to be perpendicular to $p_4$, $p_5$, and to each other.
The final expression does, however, turn out to simplify a lot by expressing the part of the result proportional to \((D_s-2)^2\) in terms of
\begin{align}
(k_1+k_2)^2 &= k_1^2 + k_2^2 + 2\bar{k}_1 \cdot \bar{k_2} - \mu_{12},
\end{align}
rather than \(\mu_{12}\), a simplification which also takes place for the other butterfly-type topologies.

The result is
\begin{align}
  &\Delta_{430}(1^+,2^+,3^+,4^+,5^+) =
  \nonumber\\&
  -\frac{i s_{12} \tr_+(1345)}{2 \A12\A23\A34\A45\A51 s_{13}}
  \big( 2\MP{k_1}{\omega_{123}}+s_{23} \big)
  \times \nonumber\\&
  \Big(
    2(D_s-2)(\mu_{11}+\mu_{22})\mu_{12}
    + (D_s-2)^2\mu_{11}\mu_{22}\frac{(k_1+k_2)^2+s_{45}}{s_{45}}
  \Big).
  \label{eq:5ptallplus_bt32}
\end{align}

\subsection{The massive double-triangle butterfly: $(330;M_2)$}

This topology can be parametrized by
\begin{align}
  \bar{k}_1^\mu &= a_1 p_1^\mu + a_2 \flm{p_{23}} + a_3 \tfrac{\A{\fl{p_{23}}}4}{2\A14}\cv1{\fl{p_{23}}} + a_4 \tfrac{\B{\fl{p_{23}}}4}{2\B14} \cv{\fl{p_{23}}}1 , \nn \\
  \bar{k}_2^\mu &= b_1 p_4^\mu + b_2 p_5^\mu + b_3 \tfrac{\A51}{2\A41}\cv45 + b_4 \tfrac{\B51}{2\B41}\cv54 ,
\end{align}
where
\begin{align}
  a_1 &= 1 \,, &
  a_2 &= 0 \,, &
  a_3 &= \tau_1 \,, &
  a_4 &= \tau_2 \,, \nn \\
  b_1 &= 0 \,, &
  b_2 &= 1 \,, &
  b_3 &= \tau_3 \,, &
  b_4 &= \tau_4 \,, \nn \\
  \mu_{11} &= \bar{k}_1^2 \,, &
  \mu_{22} &= \bar{k}_2^2 \,, &
  \mu_{12} &= s_{45} \tau_5 \,.
\end{align}

The general integrand has $146$ coefficients in terms of the ISPs
\begin{align}
(k_1 \cdot \omega_{1 \flat -}) \, , \,\; (k_2 \cdot \omega_{45-}) \, , \,\; (k_1 \cdot \omega_{1 \flat +}) \, , \,\; (k_2 \cdot \omega_{45+}) \, , \,\; \mu_{11} \, , \,\; \mu_{12} \, , \,\; \mu_{22} \,,
\end{align}
where $\omega_{1 \flat -}^{\mu}$ and $\omega_{1 \flat +}^{\mu}$ are defined to be perpendicular to $p_1$, $p_{23}$, and to each other, while $\omega_{45 -}^{\mu}$ and $\omega_{45 +}^{\mu}$ are defined as above.

The result is
\begin{align}
  &\Delta_{330;M_2}(1^+,2^+,3^+,4^+,5^+) =
  \nonumber\\&
  -\frac{i \tr_+(1345)}{2\A12\A23\A34\A45\A51}
  \frac{ s_{45}-s_{23} }{s_{13}}
  \times \nonumber\\&
  \Big(
     2(D_s-2)(\mu_{11}+\mu_{22})\mu_{12}
    + (D_s-2)^2\mu_{11}\mu_{22}\frac{(k_1+k_2)^2+s_{45}}{s_{45}}
  \Big).
  \label{eq:5ptallplus_bt22m2}
\end{align}

\subsection{The five-leg double-triangle butterfly: $(330;5L)$}

This topology can be parametrized by
\begin{align}
  \bar{k}_1^\mu &= a_1 p_1^\mu + a_2 p_2^\mu + a_3 \tfrac{\A23}{2\A13}\cv12 + a_4 \tfrac{\B23}{2\B13}\cv21 , \nn \\
  \bar{k}_2^\mu &= b_1 p_4^\mu + b_2 p_5^\mu + b_3 \tfrac{\A51}{2\A41}\cv45 + b_4 \tfrac{\B51}{2\B41}\cv54 ,
  \label{eq:oseqs-bt225L}
\end{align}
where
\begin{align}
  a_1 &= 1 \,, &
  a_2 &= 0 \,, &
  a_3 &= \tau_1 \,, &
  a_4 &= \tau_2 \,, \nn \\
  b_1 &= 0 \,, &
  b_2 &= 1 \,, &
  b_3 &= \tau_3 \,, &
  b_4 &= \tau_4 \,, \nn \\
  \mu_{11} &= \bar{k}_1^2 \,, &
  \mu_{22} &= \bar{k}_2^2 \,, &
  \mu_{12} &= s_{45} \tau_5 \,.
\end{align}

The general integrand has $146$ coefficients in terms of the ISPs
\begin{align}
(k_1 \cdot \omega_{123}) \, , \,\; (k_1 \cdot p_3) \, , \,\; (k_2 \cdot \omega_{453}) \, , \,\; (k_2 \cdot p_3) \, , \,\; \mu_{11} \, , \,\; \mu_{12} \, , \,\; \mu_{22} \,,
\end{align}
with \(\omega_{123}^{\mu}\) and \(\omega_{453}^{\mu}\) defined in
analogy with eq. (\ref{omegadef}).
We pick this basis rather than a completely spurious one, as it makes the result simplify:
\begin{align}
&  \Delta_{330;5L}(1^+ \! ,2^+, \! 3^+, \! 4^+, \! 5^+) \; = \; -\frac{i}{\A12\A23\A34\A45\A51} \times \nn \\\Bigg(
& \frac{1}{2} \left( \tr_+(1245) - \frac{\tr_+(1345) \tr_+(1235)}{s_{13} s_{35}} \right) \bigg( 2 (D_s-2) (\mu_{11} + \mu_{22}) \mu_{12} \nn \\
& \;\;\; + (D_s-2)^2 \mu_{11} \mu_{22} \frac{ 4 (k_1 \! \cdot \! p_3) (k_2 \! \cdot \! p_3) + (k_1+k_2)^2 (s_{12} + s_{45}) + s_{12}s_{45} }{s_{12}s_{45}} \bigg) \nn \\
& \; + (D_s-2)^2 \mu_{11} \mu_{22} \bigg[ \; (k_1+k_2)^2 s_{15} \nn \\
& \;\;\; + \tr_+(1235) \left( \frac{(k_1+k_2)^2}{2s_{35}} - \frac{k_1 \! \cdot \! p_3}{s_{12}} \left( 1 + \frac{2(k_2 \! \cdot \! \omega_{543})}{s_{35}} + \frac{s_{12}-s_{45}}{s_{35} s_{45}} (k_2-p_5)^2 \right) \right) \\
& \;\;\; + \tr_+(1345) \left( \frac{(k_1+k_2)^2}{2s_{13}} - \frac{k_2 \! \cdot \! p_3}{s_{45}} \left( 1 + \frac{2(k_1 \! \cdot \! \omega_{123})}{s_{13}} + \frac{s_{45}-s_{12}}{s_{12} s_{13}} (k_1-p_1)^2 \right) \right) \bigg] \Bigg). \nn
\end{align}
The two terms proportional to $(k_1-p_1)^2$ and $(k_2-p_5)^2$, which vanish on the cut, are included
in order to absorb some terms which would otherwise appear in lower point cuts.
\\ \\
In addition to those outlined above we have computed the cuts $(421)$, $(321;5L)$, $(420)$,
$(320;5L)$, and $(220;5L)$, which all vanish using the integrands presented above as subtraction
terms.

\section{The planar five gluon two-loop amplitude \label{sec:5gresult}}

Since we only have the planar primitive amplitude, we are restricted to
the leading colour part of the full amplitude. The colour structure of the two loop
amplitude can be written in terms of a decomposition into single and double traces
of $SU(N_c)$ generators. Each of these kinematic objects will have a further expansion in $N_c$
(here we use a standard notation for the permutations, see for example \cite{Bern:1994zx}):
\begin{align}
  \mathcal{A}_5 &=
  g_s^7 \sum_{\sigma \in S_5} N_c^2 \, {\rm tr}\left(T^{a_{\sigma(1)}} T^{a_{\sigma(2)}} T^{a_{\sigma(3)}} T^{a_{\sigma(4)}} T^{a_{\sigma(5)}}\right)
  \left( A^{(2)}_{5;1;1} + \frac{1}{N_c^2} A^{(2)}_{5;1;2} \right) \nonumber\\&
  + \sum_{\sigma \in S_5 / S_{5;3}} N_c \,
  {\rm tr}\left(T^{a_{\sigma(1)}} T^{a_{\sigma(2)}}\right) {\rm tr}\left(T^{a_{\sigma(3)}}
  T^{a_{\sigma(4)}} T^{a_{\sigma(5)}}\right) A^{(2)}_{5;3},
  \label{eq:colourdecomp}
\end{align}
where $T^a$ are the generators of $SU(N_c)$ and $g_s$ is the strong coupling constant.
The decomposition of the full colour amplitude into primitive amplitudes is beyond the scope of this
work since $A_{5;1;2}$ and $A_{5;3}$ will recieve contributions from planar and non-planar primitives.
The decomposition for the leading colour piece is as follows:
\begin{align}
  &\mathcal{A}_5(1^+,2^+,3^+,4^+,5^+)|_{\text{leading colour}} = \nonumber\\&
  \:\: g_s^7 N_c^2 \sum_{\sigma \in S_5} {\rm tr}\left(T^{a_{\sigma(1)}} T^{a_{\sigma(2)}} T^{a_{\sigma(3)}} T^{a_{\sigma(4)}} T^{a_{\sigma(5)}}\right)
  A_5^{(2)} \! \big(\sigma(1)^+ \!, \sigma(2)^+ \!, \sigma(3)^+ \!, \sigma(4)^+ \!, \sigma(5)^+
  \big),
  \label{eq:leading}
\end{align}
where we have chosen to abbreviate the leading colour partial amplitude to $A^{(2)}_{5} = A^{(2)}_{5;1;1}$. Its
decomposition into primitive amplitudes is:
\begin{align}
  &A_5^{(2)}(1^+,2^+,3^+,4^+,5^+) = A_5^{(2),\text{bare}}(1^+,2^+,3^+,4^+,5^+) - \frac{11}{3\eps}A_5^{(1)}(1^+,2^+,3^+,4^+,5^+), \\
  &A_5^{(2),\text{bare}}(1^+,2^+,3^+,4^+,5^+) = \sum_{i=1}^5 A_5^{[P]}(1^+,2^+,3^+;4^+,5^+),
  \label{eq:partial}
\end{align}
where we have included the appropriate UV counter-term.  The primitive amplitude $A_5^{[P]}$ can be
written in terms of the eight integral families for which the coefficients can be determined from
the cuts of the previous section.

After eliminating spurious terms which integrate to zero, the final result is
\begin{align}
& A_5^{[P]}(1^+,2^+,3^+;4^+,5^+) = \frac{i}{\A12\A23\A34\A45\A51} \Big(
    c_{431} I_{431} \left[F_1\right] \nonumber\\
  + \,&c^{T}_{431} I_{431} \left[F_1 \; (k_1+p_5)^2\right]
  + c_{331;M_1} I_{331;M_1} \left[F_1\right]
  + c_{331;M_2} I_{331;M_2} \left[F_1\right]
  + c_{331;5L}  I_{331;5L} \left[F_1\right] \nonumber\\
  + \,&c_{430} \big( s_{23} I_{430} \left[F_3 \; ((k_1+k_2)^2 +s_{45})\right]
    + I_{430} \left[F_3 \; ((k_1+k_2)^2 +s_{45}) \; 2\MP{k_1}{\omega_{123}}\right] \big) \nonumber\\
  + \,&c_{330;M_1} I_{330;M_1} \left[F_3 \; ((k_1+k_2)^2 +s_{45})\right]
  + c_{330;M_2} I_{330;M_2} \left[F_3 \; ((k_1+k_2)^2 +s_{45})\right] \nonumber\\
  + \,&c_{330;5L}^{a} I_{330;5L} \left[F_3  \; N_1(k_1, k_2, 1, 2, 3, 4, 5)\right]
  + c_{330;5L}^{b} I_{330;5L} \left[F_3  \; N_2(k_1, k_2, 1, 2, 3, 4, 5)\right] \nonumber\\
  + \,&c_{330;5L}^{c} I_{330;5L} \left[F_3  \; N_2(k_2, k_1, 5, 4, 3, 2, 1)\right]
  + c_{330;5L}^{d} I_{330;5L} \left[F_3 \; (k_1+k_2)^2 \right] \Big),
\label{eq:primitive}
\end{align}
where
\begin{equation}
  \begin{aligned}
   &\begin{aligned}
  c_{431} &= -\frac{s_{12} s_{23} s_{34} s_{45}^2 s_{15}}{\trfive} \,, &
  c_{431}^T &= -\frac{s_{12} s_{23} s_{45} \trp(1345)}{\trfive} \,, \\
  c_{331;M_1} &= -\frac{s_{34} s_{45}^2 \trp(1235)}{\trfive} \,, &
  c_{331;M_2} &= -\frac{s_{15} s_{45}^2 \trm(1234)}{\trfive} \,, \\
  c_{331;5L} &= \frac{s_{12} s_{23} s_{34} s_{45} s_{15}}{\trfive} \,, &
  c_{430} &= -\frac{s_{12} \trp(1345)}{2 s_{13} s_{45}} \,, \\
  c_{330;M_1} &= -\frac{(s_{45}-s_{12})\trp(1345)}{2 s_{13} s_{45}} \,, &
  c_{330;M_2} &= -\frac{(s_{45}-s_{23})\trp(1345)}{2 s_{13} s_{45}} \,, \\
  c_{330;5L}^b &= \frac{\trp(1235)}{2s_{35}s_{12}} \,, &
  c_{330;5L}^c &= \frac{\trp(1345)}{2s_{13}s_{45}} \,, \\
    \end{aligned} \\
    &\begin{aligned}
  c_{330;5L}^a &= -\frac{1}{2}\left(\trp(1245)-\frac{\trp(1235)\trp(1345)}{s_{13}s_{35}}\right) \,, \\
  c_{330;5L}^d &= c_{330;5L}^a \frac{s_{12}+s_{45}}{s_{12}s_{45}} - s_{12} c_{330;5L}^b - s_{45} c_{330;5L}^c - s_{15} \,,
    \end{aligned}
  \end{aligned}
  \label{eq:coeffsv2}
\end{equation}
and
\begin{equation}
  \begin{aligned}
  &F_1 = (D_s-2)(\mu_{11}\mu_{22}+\mu_{11}\mu_{33}+\mu_{22}\mu_{33}) + 4(\mu_{12}^2 - 4\mu_{11}\mu_{22}), \\
  &F_3 = (D_s-2)^2\mu_{11}\mu_{22}, \\
  &N_1(k_1, k_2, 1, 2, 3, 4, 5) = \frac{1}{s_{12}s_{45}}\left( 4\MP{k_1}{p_3}\MP{k_2}{p_3} + s_{12}s_{45} \right), \\
  &N_2(k_1, k_2, 1, 2, 3, 4, 5) = \frac{2}{s_{45}} \MP{k_1}{p_3} \left(s_{35}s_{45} - (s_{12}-s_{45})2\MP{k_2}{p_5}\right).
  \end{aligned}
  \label{eq:integralsv2}
\end{equation}
For performing the integrals over the higher-dimensional \(\mu\)-parameters, we use the Schwinger parametrization technique described in \cite{Bern:2002tk}. Defining
\begin{align}
\int \! \id \mu &=
  \int \frac{\id^{-2 \epsilon} k_1^{[-2 \epsilon]}}{(2\pi)^{-2\epsilon}}
  \int \frac{\id^{-2 \epsilon} k_2^{[-2 \epsilon]}}{(2\pi)^{-2\epsilon}},
\end{align}
to be the higher-dimensional part of the integral, the result for the specific insertions is that
\begin{align}
\label{eq:muinsertions}
\int \! \id \mu \; \big( \mu_{12}^2 - 4 \mu_{11} \mu_{22} \big) I_A^{[4-2\eps]}
    &= -2 \eps( 2 \eps + 1 ) I_A^{[6-2\eps]}, \nn \\
\int \! \id \mu \; \big( \mu_{11} \mu_{22} + \mu_{11} \mu_{33} + \mu_{22} \mu_{33} \big) I_A^{[4-2\eps]}
&= 3 \eps^2 I_A^{[6-2\eps]} + 2 \eps (\eps-1) \!\!\! \sum_{i,j \in P(A)} \!\!\! \mathbf{i}^+ \mathbf{j}^+ I_A^{[8-2\eps]}, \nn \\
\int \! \id \mu \; \mu_{11} \mu_{22} I^{[4-2\eps]}_B
    &= \eps^2 I_B^{[6-2\eps]}, \\
\int \! \id \mu \; (\mu_{11} + \mu_{22}) \mu_{12} I^{[4-2\eps]}_B &= 0, \nn
\end{align}
where $A=\{(431), (331;M_1), (331;M_2), (331;5L)\}$ and $B=\{(430), (330;M_1), (330;M_2), (330;5L)\}$.
The last identity has already been applied in eq. (\ref{eq:primitive}).  The set $P(A)$ includes all possible
ways to increase the power of any two propagators along a given branch ($k_1$, $k_2$, or $k_1 \! + \! k_2$) of a
topology $A$. Explicitly we can write the sum as,
\begin{align}
  \sum_{i,j\in P(n_1 n_2 n_{12};x)} \!\!\!\! = \;
      \sum_{i=1}^{n_1}\sum_{j=i}^{n_1} \;
    + \sum_{i=n_1+1}^{n_1+n_2}\sum_{j=i}^{n_1+n_2} \;
    + \! \sum_{i=n_1+n_2+1}^{n_1+n_2+n_{12}}\sum_{j=i}^{n_1+n_2+n_{12}} \!\! ,
  \label{eq:dblpropsum}
\end{align}
where \(n_1\) denotes the number of propagators on the \(k_1\)-branch, \(n_2\) the number of propagators on the \(k_2\)-branch, and \(n_{12}\) the number of propagators on the \(k_1 \! + \! k_2\)-branch.
For example the $(331;x)$ topologies \(I^{[8-2\eps]}_{331;x}(1,1,1,1,1,1,1)\) expand to
\begin{align}
\sum_{i,j \in P(331;x)} & \!\!\!\!\!\! \mathbf{i}^+ \mathbf{j}^+ I^{[8-2\eps]}_{331;x}(1,1,1,1,1,1,1) \; = \nonumber\\
 &I^{[8-2\eps]}_{331;x}(3,1,1,1,1,1,1)
+ I^{[8-2\eps]}_{331;x}(2,2,1,1,1,1,1)
+ I^{[8-2\eps]}_{331;x}(2,1,2,1,1,1,1)\nonumber\\
+&I^{[8-2\eps]}_{331;x}(1,3,1,1,1,1,1)
+ I^{[8-2\eps]}_{331;x}(1,2,2,1,1,1,1)
+ I^{[8-2\eps]}_{331;x}(1,1,3,1,1,1,1)\nonumber\\
+&I^{[8-2\eps]}_{331;x}(1,1,1,3,1,1,1)
+ I^{[8-2\eps]}_{331;x}(1,1,1,2,2,1,1)
+ I^{[8-2\eps]}_{331;x}(1,1,1,2,1,2,1)\nonumber\\
+&I^{[8-2\eps]}_{331;x}(1,1,1,1,3,1,1)
+ I^{[8-2\eps]}_{331;x}(1,1,1,1,2,2,1)
+ I^{[8-2\eps]}_{331;x}(1,1,1,1,1,3,1)\nonumber\\
+&I^{[8-2\eps]}_{331;x}(1,1,1,1,1,1,3).
\end{align}

\subsection{Analytical result for the butterfly-type topologies \label{sec:bowtie}}

As the integrals for the butterfly topologies are products of one-loop integrals, they can easily be
found analytically. As the six-dimensional boxes, triangles and bubbles diverge as mostly
\(1/\epsilon\), the \(\epsilon^2\) from the third identity in eqs. (\ref{eq:muinsertions}) ensures that the
result is finite. Writing the butterfly contribution to the primitive
amplitude in eq. (\ref{eq:primitive})
as
\begin{align}
& A_{5 \; \tekst{butterfly}}^{[P]}(1^+,2^+,3^+;4^+,5^+) \; = \; \frac{ i(D_s-2)^2}{\A12\A23\A34\A45\A51} \times \nn \\
\Big(
& c_{430} \big( s_{23} I_{430}^a + I_{430}^b \big) + c_{330;M_1} I_{330;M_1} + c_{330;M_2} I_{330;M_2} \nn \\
& + c_{330;5L}^a I_{330;5L}^a + c_{330;5L}^b I_{330;5L}^b + c_{330;5L}^c I_{330;5L}^c + c_{330;5L}^d I_{330;5L}^d \Big),
\end{align}
the integrals are
\begin{align}
I_{430}^a &= \frac{1}{4} + \kur{O}(\epsilon) \,, \nn \\
I_{430}^b &= \frac{\trfive}{36s_{12}} + \kur{O}(\epsilon) \,, \nn \\
I_{330;M_1} &= \frac{s_{34} - 2s_{12} + 5s_{45}}{36} + \kur{O}(\epsilon) \,, \nn \\
I_{330;M_2} &= \frac{s_{15} - 2s_{23} + 5s_{45}}{36} + \kur{O}(\epsilon) \,, \nn \\
I_{330;5L}^a &= \frac{(s_{23} - 2 s_{45}) (s_{34} + 2 s_{45}) +
 s_{12} (2 s_{34} + 17 s_{45} - 2 s_{23} - 4 s_{12})}{36 s_{12} s_{45}} + \kur{O}(\epsilon) \,, \nn \\
I_{330;5L}^b &= \frac{(2s_{35} - s_{34})(2s_{13} + s_{23})}{36} + \kur{O}(\epsilon) \,, \nn \\
I_{330;5L}^c &= \frac{(2s_{13} - s_{23})(2s_{35} + s_{34})}{36} + \kur{O}(\epsilon) \,, \\
I_{330;5L}^d &= \frac{-2s_{12} + s_{15} + s_{23} + s_{34} - 2s_{45}}{36} + \kur{O}(\epsilon) \,. \nn
\end{align}
An important cross check on this part of the amplitude comes from the cancellation of un-physical
poles in $s_{13}$ and $s_{35}$ appearing in coefficients of eqs. (\ref{eq:coeffsv2}). We find
that, after summing over the five cyclic permutations, all such poles vanish in the partial
amplitude $A_5^{(2)}(1^+,2^+,3^+,4^+,5^+)$. The finite remainder is
\begin{align}
\kur{A}_{5\;\tekst{butterfly}}^{(2)} \! &= \frac{i (D_s-2)^2}{\A12 \A23 \A34 \A45 \A51} \frac{-1}{72 \; s_{12} s_{23} s_{34} s_{45} s_{15}} \sum_{\tekst{cyclic}} X(1,2,3,4,5),
\end{align}
where $X$ can be written as
\begin{align}
X(1,2,3,4,5) &= s_{12}^2 \Big( s_{23} \big( s_{12} s_{15} s_{23} s_{34} + s_{15} s_{23}^2 s_{34} - 2 s_{15} s_{23} s_{34}^2 -
      2 s_{12} s_{15} s_{23} s_{45} + s_{12} s_{23}^2 s_{45} \nn \\
& + s_{15} s_{23}^2 s_{45} +
      s_{12} s_{23} s_{34} s_{45} + 12 s_{15} s_{23} s_{34} s_{45} - 2 s_{23}^2 s_{34} s_{45} +
      2 s_{15} s_{34}^2 s_{45} \nn \\
& - 2 s_{23} s_{34}^2 s_{45} + 2 s_{23} s_{34} s_{45}^2 +
      2 s_{34}^2 s_{45}^2 \big) - \big( s_{23}^2 s_{45} +
      s_{15} s_{34} s_{45} \nn \\
& + s_{23} s_{34} s_{45} - s_{15} s_{23} s_{34} - s_{15} s_{23} s_{45} \big) \trfive \Big).
\end{align}

\section{Numerical evaluation \label{sec:5gnum}}

Three of the integrals appearing in the amplitude in eq. (\ref{eq:primitive}) have five scales and are yet unknown
analytically. We therefore opt for a numerical evaluation in order to check the universal
infra-red pole structure:
\begin{align}
  A_5^{(2)} \! (1^+\!,2^+\!,3^+\!,4^+\!,5^+) = -\left(\frac{1}{\eps^2}\sum_{i=1}^{5} \left(
  \frac{\mu_R^2}{-s_{i \; i+1}}
  \right)^{\! \eps} + \frac{11}{6\eps} \right)\! A_5^{(1)} \! (1^+\!,2^+\!,3^+\!,4^+\!,5^+) +
  \mathcal{O}(\eps^0),
  \label{eq:IRpoles}
\end{align}
where the $D$-dimensional one-loop amplitude is given by \cite{Bern:1995db,Bern:1996ja},
\begin{align}
  A_5^{(1)}&(1^+\!,2^+\!,3^+\!,4^+\!,5^+) = \frac{-i\eps(1-\eps)}{\A12\A23\A34\A45\A51}\Big(
    s_{12} s_{23} I^{8-2\eps}_{4;1234}[1]
  + s_{23} s_{34} I^{8-2\eps}_{4;2345}[1]\nonumber\\&
  + s_{34} s_{45} I^{8-2\eps}_{4;3451}[1]
  + s_{45} s_{15} I^{8-2\eps}_{4;4512}[1]
  + s_{15} s_{12} I^{8-2\eps}_{4;5123}[1]
  + 2(2-\eps) \trfive I^{10-2\eps}_{5;12345}[1]
  \Big).
  \label{eq:1l5ptamp}
\end{align}
We use two techniques for the numerical integration. Firstly we use the Mellin-Barnes approach with
the help of {\sc AMBRE}~\cite{Gluza:2007rt}\footnote{We thank Tristan Dennen for providing a copy of his private
implementation based on the AMBRE algorithm.}, M.~Czakon's {\tt MB.m} \cite{Czakon:2005rk},
A.~V.~Smirnov's {\tt MBresolve.m} \cite{Smirnov:2009up} and D.~A.~Kosower's {\tt barnesroutines.m}.
The second approach uses sector decomposition via the {\sc FIESTA} Mathematica package
\cite{Smirnov:2008py,Smirnov:2009pb}. The results of the numerical evaluation are shown in table
\ref{tab:ampeval} using the phase-space point:
\begin{align}
\begin{aligned}
  p_1 &= (8/3, 1/2, i/2, 8/3) \,, \\
  p_2 &= (0, 1/2, -i/2, 0) \,, \\
  p_3 &= (-1, 1, 2i, 2) \,, \\
  p_4 &= (61.23163693, -59.08662701, 76.08662701i, 77.76412206) \,, \\
  p_5 &= (-62.89830359, 57.08662701, -78.08662701i, -82.43078872) \,.
  \label{eq:pspoint}
\end{aligned}
\end{align}
Though the configuration is complex, it has been constructed so that the
exact kinematics can be obtained using the following values of the invariants:
\begin{align}
  s_{12} &= -1 \,, &
  s_{23} &= -3 \,, &
  s_{34} &= -11 \,, &
  s_{45} &= -19 \,, &
  s_{15} &= -31 \,, &
  \trfive &= -\sqrt{154429} \,.
  \label{eq:psinvariants}
\end{align}
\begin{table}
  \centering
  \begin{tabular}{|c|c|c|c|}
    \hline
    & $\eps^{-2}$ & $\eps^{-1}$ & $\eps^{0}$ \\
    \hline
    $A_5^{[P]}(1,2,3;4,5)$ & $-145.03 \pm 0.01$ & $473.37 \pm 0.10$ & $-1643.16 \pm 0.60$ \\
    \hline
    $A_5^{[P]}(2,3,4;5,1)$ & $-23.00 \pm 0.00$ & $86.54 \pm 0.02$ & $-229.22 \pm 0.09$ \\
    \hline
    $A_5^{[P]}(3,4,5;1,2)$ & $-70.65 \pm 0.00$ & $118.03 \pm 0.02$ & $3279.84 \pm 0.10$ \\
    \hline
    $A_5^{[P]}(4,5,1;2,3)$ & $5.19 \pm 0.00$ & $-15.11 \pm 0.00$ & $45.91 \pm 0.01$ \\
    \hline
    $A_5^{[P]}(5,1,2;3,4)$ & $-159.87 \pm 0.01$ & $625.73 \pm 0.10$ & $-794.94 \pm 0.90$ \\
    \hline
    $A_5^{(2),\text{bare}}(1,2,3,4,5)$ & $-393.36 \pm 0.02$ & $1288.56 \pm 0.20$ & $658.43 \pm 1.00$ \\
    \hline
    $ \mathcal{I}_5^{(1),\text{bare}} A_5^{(1)}(1,2,3,4,5)$ & $-393.35 \pm 0.02$ & $1288.50 \pm 0.08$ & $-2627.61 \pm 0.20$ \\
    \hline
  \end{tabular}
  \caption{Numerical values for the poles and finite part of the amplitude at the example
  phase-space point in eq. (\ref{eq:pspoint}). The loop integrals have been evaluated with a
  pre-factor of $-(4\pi)^{-D} e^{-2\eps \gamma_E}$ removed. We have also
  stripped out the overall helicity factor $i/(\langle 12 \rangle \langle 23
  \rangle  \langle 34 \rangle \langle 45 \rangle  \langle 51 \rangle)
  $. The final entry corresponds to the right
  hand side of eq. \eqref{eq:IRpoles} but with UV pole of $\tfrac{11}{3\eps}$ removed. Though we
  quote the sector decomposition numerical errors on the $\tfrac{1}{\eps^2}$ poles these results have also
  been computed analytically.}
  \label{tab:ampeval}
\end{table}
We can see from the values in table \ref{tab:ampeval} that the two-loop amplitude is in agreement
with the IR pole structure within the numerical integration errors.

\section{Conclusions}

In this paper we have described how $D$-dimensional integrand reduction
and generalized unitarity cuts are efficient methods for two-loop amplitude computations.

The methods presented provide a general algorithm for the reduction of any loop amplitude.  Though
the procedure does not change a lot from the four-dimensional case already presented
\cite{Mastrolia:2011pr,Kosower:2011ty,Badger:2012dp,Zhang:2012ce}, we find that the $D$-dimensional
case resolves some difficulties that can occur otherwise. We have shown that all ideals defined by
the propagators are \textit{radical ideals}. The first consequence of this is that all ideals are
\textit{prime ideals} and that each set of propagators admits only a single branch of the solution
to the on-shell constraints. We have also shown that the dimension of each ideal is $11-P$ for a
$P$ propagator system, a condition which is not always satisfied in four dimensions.

As a non-trivial application of the method, we have computed the first five-point two-loop amplitude in
QCD: the planar all-plus helicity amplitude. The final result obtained has a remarkably simple
and compact form. We learned important lessons about the benefits of choosing a good basis of ISPs
for the integrand: Firstly, we required the four dimensional limit of the integrand basis to be
manifest, a condition which is not satisfied when using a standard ordering for the polynomial
division. Secondly, we found that correctly identifying spurious directions in the bow-tie
topologies led to significant simplification in the final coefficients. Following these guidelines
we were able to find an integrand representation with only six or higher propagator
topologies with all remaining cuts evaluating to zero. As a result, we did not require
integration-by-parts to simplify the computation at any stage, since the all-plus helicity
configuration is already extremely simple at the integrand level.

A particularly interesting feature of the amplitude is the close relation to the known result in
$\mathcal{N}=4$ super Yang-Mills \cite{Bern:2006vw,Carrasco:2011mn}. At one-loop it is known that there is a close
relation between MHV amplitudes in $\mathcal{N}=4$ super Yang-Mills and self-dual Yang-Mills \cite{Bern:1996ja}:
\begin{align}
  A_n^{(1),4-2\eps}(1^+,\cdots,n^+) = -2 (4\pi)^2 \eps(1-\eps) \A12^{-4} A^{(1),[N=4],8-2\eps}(1^-,2^-,3^+,\cdots,n^+).
\end{align}
At the integrand level this corresponds to,
\begin{align}
  \Delta_n^{(1)}(1^+,\cdots,n^+) = \A12^{-4} (D_s-2)\mu_{11}^2
  \Delta_n^{(1),[N=4]}(1^-,2^-,3^+,\cdots,n^+).
\end{align}
Though the relation is not so simple at two-loops, we observe for $n=4,5$,
\begin{align}
  \Delta_n^{(2)}&(1^+,\cdots,n^+) = \nonumber\\&
  F_1 \A12^{-4} \Delta^{(2),[N=4]}(1^-,2^-,3^+,\cdots,n^+) + \text{butterfly topologies}.
\end{align}
where $F_1 = (D_s-2)\left( \mu_{11}\mu_{22} + \mu_{11}\mu_{33} + \mu_{22}\mu_{33}  \right) + 4\left(
\mu_{12}^2 - 4\mu_{11}\mu_{22} \right)$ as given in eq. \eqref{eq:F1def}. As noted in section
\ref{sec:bowtie}, all the remaining butterfly contributions are finite, though we clearly see from eqs.
\eqref{eq:muinsertions} that the dimension shifting formula no longer holds. It would
certainly be interesting if there was a better way of understanding this rather unexpected
connection between the two theories.

The main prospects for future research would be to complete the computation of the other helicity
configurations for $5$-gluon scattering. We would like to investigate extensions of the integrand
reduction method to the non-planar case, where some recent developments using the colour-kinematics
duality may be extremely helpful \cite{Bern:2008qj,Bern:2013yya}.

\acknowledgments{%
We are grateful to Zvi Bern, Pierpaolo Mastrolia, Simon Caron-Huot, Pierre Vanhove, and Rijun Huang
for useful discussions. We would especially like to thank Tristan Dennen for his advice on numerical
integration techniques and for careful reading of the manuscript. We also thank Donal O'Connell for clarifications of the six-dimensional
spinor-helicity formalism. The work of YZ is supported by Danish Council for Independent Research
(FNU) grant 11-107241. This work was also supported by the Research Executive Agency (REA) of the European Union
through the Initial Training Network LHCPhenoNet under contract PITN-GA-2010-264564%
}

\appendix
\section{Momentum twistor parametrization of the kinematics \label{app:momtwistor}}

An analytical calculation of a scattering amplitude usually results in a huge function of the scalar products $s_{ij}$, the
antisymmetric product $\trfive$ defined in eq. (\ref{trfivedef}), and spinor products $\langle ij \rangle$ and
$[ij]$. These quantities are not independent, because of energy-momentum conservation and algebraic constraints like
the Schouten identity, but to automate the reduction to some minimal set is hard beyond \(2 \rightarrow 2\) kinematics.

In this paper, we use the {\it momentum twistor
  parametrization} for the analytical computations, in which the kinematics can be represented by momentum twistors
$Z_i(\lambda_i, \mu_i)$ for each momentum
\cite{Hodges:2009hk}.

The advantage of momentum twistor
variables is that all identities like the Schouten identity,
energy-momentum conservation, etc. are satisfied automatically, making the final simplification process
straightforward. It is also easy to convert the momentum-twistor
variables back to the traditional kinematic variables.

The first two components $\lambda_i$, of a momentum twistor $Z_i(\lambda_i,
\mu_i)$ are the holomorphic spinors, whereas the anti-holomorphic spinors are obtained via
\begin{align}
  \tilde{\lambda}_i =
  \frac{\A{i,}{\,i+1}\mu_{i-1} + \A{i+1,}{\,i-1}\mu_i + \A{i-1,}{\,i}\mu_{i+1}}
  {\A{i,}{\,i+1}\A{i-1,}{\,i}},
  \label{eq:momtwistoreq}
\end{align}
where $\A{i}{j}$ denotes the usual spinor products.

We may use the
symmetries of $Z$ to reduce the number of independent kinematic
quantities. The momentum twistor $Z_i(\lambda_i,
\mu_i)$ has the following symmetries:
\begin{itemize}
  \item Poincar\'{e} symmetry.
  \item $U(1)$ symmetry for each particle:    $\lambda_i \to e^{i
      \theta_i} \lambda_i$, $\mu_i \to e^{i
      \theta_i} \mu_i$.
\end{itemize}
For a $n$-particle process, there will be $4n$ momentum twistor components, but the symmetries reduce this number to $4n-10-n=3n-10$ free
components. In practice many different ways of choosing those free components are available.

\subsection{Four-point momentum twistors}
The four-point kinematics are very simple, so as a warm-up
example we will derive the four-point momentum twistors.

We choose the momentum-twistor parametrization to be,
\begin{align}
  Z =\begin{pmatrix}
\lambda_1 & \lambda_2 & \lambda_3 & \lambda_4 \\
\mu_1 & \mu_2 &\mu_3 &\mu_4
\end{pmatrix}
=
  \begin{pmatrix}
    1 & 0 & -\tfrac{1}{s} & \; -\tfrac{1}{s} \! - \! \tfrac{1}{t} \\
    0 & 1 & 1 & 1 \\
    0 & 0 & 1 & 0 \\
    0 & 0 & 0 & 1
  \end{pmatrix}.
  \label{eq:momtwistor4pt}
\end{align}
For this choice all,
the spinors $\lambda_i$ and $\tilde \lambda_i$ and the momenta $p_i$
are rational functions of the independent {\it momentum-twistor variables} $s$ and
$t$.

In this formalism $s_{12}=\langle 12 \rangle [21]=s$, and $s_{14}=\langle 14 \rangle [41]=t$, so the
twistor variables in the four-point case are just the Mandelstam variables. Any physical expression
without an overall helicity factor is calculable directly using twistors. As a simple example, the
ratio $(\langle 12\rangle \langle 34\rangle)/(\langle 13 \rangle \langle 24 \rangle)$ is
helicity-free, so by inserting eq. (\ref{eq:momtwistoreq})
and (\ref{eq:momtwistor4pt}) we get that $\langle 12 \rangle\to-1$, $\langle 34 \rangle\to-1/t$,
$\langle 13 \rangle\to-1$, and $\langle 24 \rangle \to-1/s-1/t$, and thus
\begin{eqnarray}
  \label{eq:12}
  \frac{\langle 12\rangle \langle 34\rangle}{\langle 13 \rangle
    \langle 24 \rangle}=\frac{(-1)(-1/t)}{(-1)(-1/s-1/t)}=\frac{s}{s+t}.
\end{eqnarray}
Note that this calculation is straightforward, without invoking energy-momentum conservation or the
Schouten identity, as these constraints are imposed automatically by the twistor formalism.

\subsection{Five-point momentum twistors}
There are five free components in five-point momentum twistors. The explicit form of five parameters is not unique, and it is hard to say whether
or not there is an ideal choice. One practical version is
\begin{align}
  Z =\begin{pmatrix}
\lambda_1 & \lambda_2 & \lambda_3 & \lambda_4 &\lambda_5\\
\mu_1 & \mu_2 &\mu_3 &\mu_4&\mu_5
\end{pmatrix}
=
  \begin{pmatrix}
    1 & 0 & \tfrac{1}{x_1} & \tfrac{1}{x_1} \! + \! \tfrac{1}{x_2} & \tfrac{1}{x_1} \! + \! \tfrac{1}{x_2} \! + \! \tfrac{1}{x_3} \\
    0 & 1 & 1 & 1 & 1 \\
    0 & 0 & 0 & x_4 & 1 \\
    0 & 0 & 1 & 1 & \frac{x_5}{x_4}
  \end{pmatrix}.
  \label{eq:momtwistor2}
\end{align}
Again, all the spinors $\lambda_i$ and $\tilde \lambda_i$ and the momenta $p_i$
are rational functions of the independent {\it momentum-twistor variables} $x_1$,
$x_2$, $x_3$, $x_4$ and $x_5$.

The physical five-point amplitude, after striping out the overall helicity
factor, should, by Lorentz symmetry, be a function of kinematic invariants $s_{ij}$ and
$\trfive$ only.
\begin{eqnarray}
  \label{eq:1}
  A(1,2,3,4,5)=h\cdot f(s_{ij}, \tr_5) , 
\end{eqnarray}
where $h$ contains all the helicity information of the amplitude. In
practice, if the corresponding tree amplitude is non-zero, we can
choose $h= A_\text{tree}(1,2,3,4,5)$. For the all-plus amplitude, for which the tree amplitude vanishes, we
can simply choose $h=1/(\langle 12 \rangle \langle 23 \rangle \langle
34 \rangle \langle 45 \rangle \langle 51 \rangle)$. Only five of the $s_{ij}$-variables are independent, and we choose to pick $s_{12}$, $s_{23}$,
$s_{34}$, $s_{45}$, and $s_{15}$.
Note that $\tr_5$ is not
completely independent, as
\begin{equation}
  (\tr_5)^2=\det G \begin{pmatrix}
p_1 & p_2 & p_3 & p_4 \\
p_1 & p_2 & p_3 & p_4
 \end{pmatrix}\equiv G_4( s_{12}, s_{23}, s_{34}, s_{45}, s_{15}).
 \label{trfivesq}
\end{equation}
These variables can be written as functions of the momentum-twistor variables
\begin{eqnarray}
  \label{eq:4}
s_{12}&=& x_1 , \nonumber\\
s_{23}&=&x_2 x_4 , \nonumber\\
s_{34}&=&\frac{x_1 \left(x_3 \left(x_4-1\right)+x_2 x_4\right)+x_2 x_3
  \left(x_4-x_5\right)}{x_2} , \nonumber\\
s_{45}&=&x_2
   \left(x_4-x_5\right) , \nonumber\\
s_{15}&=&-x_3 \left(x_5-1\right) , \nonumber\\
\tr_5&=&x_1 \left(x_3 \left(x_4 \left(x_5-2\right)+1\right)+x_2 x_4
   \left(x_5-1\right)\right)+x_2 x_3 \left(x_5-x_4\right) ,
\label{s2x}
\end{eqnarray}
and inversely, the momentum twistor variables can be uniquely expressed as rational
functions of $s_{ij}$ and $\tr_5$:
\begin{eqnarray}
x_1&=&s_{12}, \nonumber\\
x_2&=&\frac{s_{12} \left(s_{23}-s_{15}\right)+s_{23} s_{34}+s_{15} s_{45}-s_{34} s_{45}-\tr_5}{2
   s_{34}}, \nonumber\\
x_3&=&\frac{\left(s_{23}-s_{45}\right) \left(s_{23} s_{34}+s_{15} s_{45}-s_{34}
   s_{45}-\tr_5\right)+s_{12} \left(s_{15}-s_{23}\right) s_{23}+s_{12} \left(s_{15}+s_{23}\right)
   s_{45}}{2 \left(s_{12}+s_{23}-s_{45}\right) s_{45}}, \nonumber \\
x_4&=&-\frac{s_{12} \left(s_{23}-s_{15}\right)+s_{23} s_{34}+s_{15} s_{45}-s_{34}
   s_{45}+\tr_5}{2 s_{12} \left(s_{15}-s_{23}+s_{45}\right)}, \nonumber\\
x_5&=&\frac{\left(s_{23}-s_{45}\right) \left(s_{12}
   \left(s_{23}-s_{15}\right)+s_{23} s_{34}+s_{15} s_{45}-s_{34} s_{45}+\tr_5\right)}{2 s_{12} s_{23}
   \left(-s_{15}+s_{23}-s_{45}\right)}.
\label{x2s}
\end{eqnarray}
Mathematically, the kinematic space of five-particle massless
scattering is defined by the {\it affine variety}
$V_5=\mathcal Z(\tr_5^2-G_4( s_{12}, s_{23}, s_{34}, s_{45},
s_{15}))$ in $\mathbb C^6$, spanned by the variables $s_{12}$, $s_{23}$,
$s_{34}$, $s_{45}$, $s_{15}$, and $\tr_5$. This affine variety is {\it birationally
equivalent} to $\mathbb C^5$ spanned by $x_1$,
$x_2$, $x_3$, $x_4$ and $x_5$ via the maps (\ref{s2x}) and
(\ref{x2s}).

Note that momentum-twistor variables $x_1$,
$x_2$, $x_3$, $x_4$ and $x_5$, are only equivalent to the kinematics
variables $s_{ij}$ and $tr_5$, but not to the spinor products $\langle ij
\rangle$ or $[ij]$, as they contain the
additional phase information.

Hence our strategy for calculations of five-point massless amplitudes using
the momentum-twistor formalism can be summarized as follows:
\begin{itemize}
\item Calculate the amplitude $A(1,2,3,4,5)$ in  momentum-twistor
  variables and simplify the result.
\item Isolate the helicity factor $h$, as $A(1,2,3,4,5)=h \cdot f$.
\item Use eq. (\ref{x2s}) to rewrite $f(x_1, x_2, x_3, x_4, x_5)$ as
  $f(s_{ij},\tr_5)$, where eq. (\ref{trfivesq}) ensures us that the result can be expressed as mostly linear in \(\trfive\).
\end{itemize}


\section{Feynman rules for gluons and scalars \label{feynmanappendix}}

The tree-level amplitudes in this paper have been computed using the following colour ordered Feynman rules,
\begin{align}
\tekst{prop}_{g}^{\mu \nu} &= \frac{-ig^{\mu \nu}}{p^2}, & \tekst{prop}_{s} &= \frac{i}{p^2},
\end{align}
\begin{align}
V^{\mu_1 \mu_2 \mu_3}_{ggg} &= \frac{i}{\sqrt{2}} \Big( g^{\mu_2 \mu_3} (p_2 - p_3)^{\mu_1} + g^{\mu_3 \mu_1} (p_3 - p_1)^{\mu_2} + g^{\mu_1 \mu_2} (p_1 - p_2)^{\mu_3} \Big),
\end{align}
\begin{align}
V^{\mu_1 \mu_2 \mu_3 \mu_4}_{gggg} &= i g^{\mu_1 \mu_3} g^{\mu_2 \mu_4} - \frac{i}{2} \big( g^{\mu_1 \mu_2} g^{\mu_3 \mu_4} + g^{\mu_4 \mu_1} g^{\mu_2 \mu_3} \big),
\end{align}
\begin{align}
V^{\mu}_{sgs} &= \frac{i}{\sqrt{2}} \big( p_1 - p_3 )^{\mu},
\end{align}
\begin{align}
V^{\mu \nu}_{ggss} &= \frac{i}{2} g^{\mu \nu} \; , & V^{\mu \nu}_{gsgs} &= -i g^{\mu \nu} \; ,\nn \\
V_{sss's'} &= -\frac{i}{2} \; , & V_{ss'ss'} &= i \; ,
\label{scalarfourpoint}
\end{align}
where the subscribed $g$ refers to gluons, and where the scalars $s$ and $s'$ may have different
flavours. The rules in this appendix are consistent with those listed in \cite{Dixon:1996wi}.


\begin{thebibliography}{10}

\bibitem{Bern:1994zx}
Z.~Bern, L.~J. Dixon, D.~C. Dunbar, and D.~A. Kosower, {\it {One loop n point
  gauge theory amplitudes, unitarity and collinear limits}},  {\em Nucl.Phys.}
  {\bf B425} (1994) 217--260,
  [\href{http://xxx.lanl.gov/abs/hep-ph/9403226}{{\tt hep-ph/9403226}}].

\bibitem{Bern:1994cg}
Z.~Bern, L.~J. Dixon, D.~C. Dunbar, and D.~A. Kosower, {\it {Fusing gauge
  theory tree amplitudes into loop amplitudes}},  {\em Nucl.Phys.} {\bf B435}
  (1995) 59--101, [\href{http://xxx.lanl.gov/abs/hep-ph/9409265}{{\tt
  hep-ph/9409265}}].

\bibitem{Britto:2004ap}
R.~Britto, F.~Cachazo, and B.~Feng, {\it {New recursion relations for tree
  amplitudes of gluons}},  {\em Nucl.Phys.} {\bf B715} (2005) 499--522,
  [\href{http://xxx.lanl.gov/abs/hep-th/0412308}{{\tt hep-th/0412308}}].

\bibitem{Britto:2004nc}
R.~Britto, F.~Cachazo, and B.~Feng, {\it {Generalized unitarity and one-loop
  amplitudes in N=4 super-Yang-Mills}},  {\em Nucl.Phys.} {\bf B725} (2005)
  275--305, [\href{http://xxx.lanl.gov/abs/hep-th/0412103}{{\tt
  hep-th/0412103}}].

\bibitem{Ellis:2007br}
R.~Ellis, W.~Giele, and Z.~Kunszt, {\it {A Numerical Unitarity Formalism for
  Evaluating One-Loop Amplitudes}},  {\em JHEP} {\bf 0803} (2008) 003,
  [\href{http://xxx.lanl.gov/abs/0708.2398}{{\tt arXiv:0708.2398}}].

\bibitem{Forde:2007mi}
D.~Forde, {\it {Direct extraction of one-loop integral coefficients}},  {\em
  Phys.Rev.} {\bf D75} (2007) 125019,
  [\href{http://xxx.lanl.gov/abs/0704.1835}{{\tt arXiv:0704.1835}}].

\bibitem{Giele:2008ve}
W.~T. Giele, Z.~Kunszt, and K.~Melnikov, {\it {Full one-loop amplitudes from
  tree amplitudes}},  {\em JHEP} {\bf 0804} (2008) 049,
  [\href{http://xxx.lanl.gov/abs/0801.2237}{{\tt arXiv:0801.2237}}].

\bibitem{Badger:2008cm}
S.~Badger, {\it {Direct Extraction Of One Loop Rational Terms}},  {\em JHEP}
  {\bf 0901} (2009) 049, [\href{http://xxx.lanl.gov/abs/0806.4600}{{\tt
  arXiv:0806.4600}}].

\bibitem{Ossola:2006us}
G.~Ossola, C.~G. Papadopoulos, and R.~Pittau, {\it {Reducing full one-loop
  amplitudes to scalar integrals at the integrand level}},  {\em Nucl.Phys.}
  {\bf B763} (2007) 147--169,
  [\href{http://xxx.lanl.gov/abs/hep-ph/0609007}{{\tt hep-ph/0609007}}].

\bibitem{Ossola:2007ax}
G.~Ossola, C.~G. Papadopoulos, and R.~Pittau, {\it {CutTools: A Program
  implementing the OPP reduction method to compute one-loop amplitudes}},  {\em
  JHEP} {\bf 0803} (2008) 042, [\href{http://xxx.lanl.gov/abs/0711.3596}{{\tt
  arXiv:0711.3596}}].

\bibitem{Berger:2008sj}
C.~Berger, Z.~Bern, L.~Dixon, F.~Febres~Cordero, D.~Forde, {\em et.~al.}, {\it
  {An Automated Implementation of On-Shell Methods for One-Loop Amplitudes}},
  {\em Phys.Rev.} {\bf D78} (2008) 036003,
  [\href{http://xxx.lanl.gov/abs/0803.4180}{{\tt arXiv:0803.4180}}].

\bibitem{Giele:2008bc}
W.~Giele and G.~Zanderighi, {\it {On the Numerical Evaluation of One-Loop
  Amplitudes: The Gluonic Case}},  {\em JHEP} {\bf 0806} (2008) 038,
  [\href{http://xxx.lanl.gov/abs/0805.2152}{{\tt arXiv:0805.2152}}].

\bibitem{Ellis:2008qc}
R.~Ellis, W.~Giele, Z.~Kunszt, K.~Melnikov, and G.~Zanderighi, {\it {One-loop
  amplitudes for $W+$ 3 jet production in hadron collisions}},  {\em JHEP} {\bf
  0901} (2009) 012, [\href{http://xxx.lanl.gov/abs/0810.2762}{{\tt
  arXiv:0810.2762}}].

\bibitem{Mastrolia:2010nb}
P.~Mastrolia, G.~Ossola, T.~Reiter, and F.~Tramontano, {\it {Scattering
  AMplitudes from Unitarity-based Reduction Algorithm at the Integrand-level}},
   {\em JHEP} {\bf 1008} (2010) 080,
  [\href{http://xxx.lanl.gov/abs/1006.0710}{{\tt arXiv:1006.0710}}].

\bibitem{Badger:2010nx}
S.~Badger, B.~Biedermann, and P.~Uwer, {\it {NGluon: A Package to Calculate
  One-loop Multi-gluon Amplitudes}},  {\em Comput.Phys.Commun.} {\bf 182}
  (2011) 1674--1692, [\href{http://xxx.lanl.gov/abs/1011.2900}{{\tt
  arXiv:1011.2900}}].

\bibitem{Hirschi:2011pa}
V.~Hirschi, R.~Frederix, S.~Frixione, M.~V. Garzelli, F.~Maltoni, {\em
  et.~al.}, {\it {Automation of one-loop QCD corrections}},  {\em JHEP} {\bf
  1105} (2011) 044, [\href{http://xxx.lanl.gov/abs/1103.0621}{{\tt
  arXiv:1103.0621}}].

\bibitem{Bevilacqua:2011xh}
G.~Bevilacqua, M.~Czakon, M.~Garzelli, A.~van Hameren, A.~Kardos, {\em
  et.~al.}, {\it {HELAC-NLO}},  \href{http://xxx.lanl.gov/abs/1110.1499}{{\tt
  arXiv:1110.1499}}.

\bibitem{Cullen:2011xs}
G.~Cullen, N.~Greiner, G.~Heinrich, G.~Luisoni, P.~Mastrolia, {\em et.~al.},
  {\it {GoSam: A program for automated one-loop Calculations}},
  \href{http://xxx.lanl.gov/abs/1111.6534}{{\tt arXiv:1111.6534}}.

\bibitem{Badger:2012pg}
S.~Badger, B.~Biedermann, P.~Uwer, and V.~Yundin, {\it {Numerical evaluation of
  virtual corrections to multi-jet production in massless QCD}},  {\em
  Comput.Phys.Commun.} {\bf 184} (2013) 1981--1998,
  [\href{http://xxx.lanl.gov/abs/1209.0100}{{\tt arXiv:1209.0100}}].

\bibitem{Bern:1997nh}
Z.~Bern, J.~Rozowsky, and B.~Yan, {\it {Two loop four gluon amplitudes in N=4
  superYang-Mills}},  {\em Phys.Lett.} {\bf B401} (1997) 273--282,
  [\href{http://xxx.lanl.gov/abs/hep-ph/9702424}{{\tt hep-ph/9702424}}].

\bibitem{Bern:2005iz}
Z.~Bern, L.~J. Dixon, and V.~A. Smirnov, {\it {Iteration of planar amplitudes
  in maximally supersymmetric Yang-Mills theory at three loops and beyond}},
  {\em Phys.Rev.} {\bf D72} (2005) 085001,
  [\href{http://xxx.lanl.gov/abs/hep-th/0505205}{{\tt hep-th/0505205}}].

\bibitem{Bern:2006ew}
Z.~Bern, M.~Czakon, L.~J. Dixon, D.~A. Kosower, and V.~A. Smirnov, {\it {The
  Four-Loop Planar Amplitude and Cusp Anomalous Dimension in Maximally
  Supersymmetric Yang-Mills Theory}},  {\em Phys.Rev.} {\bf D75} (2007) 085010,
  [\href{http://xxx.lanl.gov/abs/hep-th/0610248}{{\tt hep-th/0610248}}].

\bibitem{Bern:2007ct}
Z.~Bern, J.~Carrasco, H.~Johansson, and D.~Kosower, {\it {Maximally
  supersymmetric planar Yang-Mills amplitudes at five loops}},  {\em Phys.Rev.}
  {\bf D76} (2007) 125020, [\href{http://xxx.lanl.gov/abs/0705.1864}{{\tt
  arXiv:0705.1864}}].

\bibitem{ArkaniHamed:2010kv}
N.~Arkani-Hamed, J.~L. Bourjaily, F.~Cachazo, S.~Caron-Huot, and J.~Trnka, {\it
  {The All-Loop Integrand For Scattering Amplitudes in Planar N=4 SYM}},  {\em
  JHEP} {\bf 1101} (2011) 041, [\href{http://xxx.lanl.gov/abs/1008.2958}{{\tt
  arXiv:1008.2958}}].

\bibitem{Chetyrkin:1981qh}
K.~Chetyrkin and F.~Tkachov, {\it {Integration by Parts: The Algorithm to
  Calculate beta Functions in 4 Loops}},  {\em Nucl.Phys.} {\bf B192} (1981)
  159--204.

\bibitem{Anastasiou:2000kg}
C.~Anastasiou, E.~Glover, C.~Oleari, and M.~Tejeda-Yeomans, {\it {Two-loop QCD
  corrections to the scattering of massless distinct quarks}},  {\em
  Nucl.Phys.} {\bf B601} (2001) 318--340,
  [\href{http://xxx.lanl.gov/abs/hep-ph/0010212}{{\tt hep-ph/0010212}}].

\bibitem{Anastasiou:2000ue}
C.~Anastasiou, E.~Glover, C.~Oleari, and M.~Tejeda-Yeomans, {\it {Two loop QCD
  corrections to massless identical quark scattering}},  {\em Nucl.Phys.} {\bf
  B601} (2001) 341--360, [\href{http://xxx.lanl.gov/abs/hep-ph/0011094}{{\tt
  hep-ph/0011094}}].

\bibitem{Anastasiou:2001sv}
C.~Anastasiou, E.~Glover, C.~Oleari, and M.~Tejeda-Yeomans, {\it {Two loop QCD
  corrections to massless quark gluon scattering}},  {\em Nucl.Phys.} {\bf
  B605} (2001) 486--516, [\href{http://xxx.lanl.gov/abs/hep-ph/0101304}{{\tt
  hep-ph/0101304}}].

\bibitem{Glover:2001af}
E.~Glover, C.~Oleari, and M.~Tejeda-Yeomans, {\it {Two loop QCD corrections to
  gluon-gluon scattering}},  {\em Nucl.Phys.} {\bf B605} (2001) 467--485,
  [\href{http://xxx.lanl.gov/abs/hep-ph/0102201}{{\tt hep-ph/0102201}}].

\bibitem{Garland:2001tf}
L.~Garland, T.~Gehrmann, E.~N. Glover, A.~Koukoutsakis, and E.~Remiddi, {\it
  {The Two loop QCD matrix element for e+ e- $\to$ 3 jets}},  {\em Nucl.Phys.}
  {\bf B627} (2002) 107--188,
  [\href{http://xxx.lanl.gov/abs/hep-ph/0112081}{{\tt hep-ph/0112081}}].

\bibitem{Garland:2002ak}
L.~Garland, T.~Gehrmann, E.~N. Glover, A.~Koukoutsakis, and E.~Remiddi, {\it
  {Two loop QCD helicity amplitudes for e+ e- $\to$ three jets}},  {\em
  Nucl.Phys.} {\bf B642} (2002) 227--262,
  [\href{http://xxx.lanl.gov/abs/hep-ph/0206067}{{\tt hep-ph/0206067}}].

\bibitem{Gehrmann:2011aa}
T.~Gehrmann, M.~Jaquier, E.~Glover, and A.~Koukoutsakis, {\it {Two-Loop QCD
  Corrections to the Helicity Amplitudes for $H \to$ 3 partons}},  {\em JHEP}
  {\bf 1202} (2012) 056, [\href{http://xxx.lanl.gov/abs/1112.3554}{{\tt
  arXiv:1112.3554}}].

\bibitem{Bern:2000dn}
Z.~Bern, L.~J. Dixon, and D.~Kosower, {\it {A Two loop four gluon helicity
  amplitude in QCD}},  {\em JHEP} {\bf 0001} (2000) 027,
  [\href{http://xxx.lanl.gov/abs/hep-ph/0001001}{{\tt hep-ph/0001001}}].

\bibitem{Bern:2000ie}
Z.~Bern, L.~J. Dixon, and A.~Ghinculov, {\it {Two loop correction to Bhabha
  scattering}},  {\em Phys.Rev.} {\bf D63} (2001) 053007,
  [\href{http://xxx.lanl.gov/abs/hep-ph/0010075}{{\tt hep-ph/0010075}}].

\bibitem{Bern:2001df}
Z.~Bern, A.~De~Freitas, and L.~J. Dixon, {\it {Two loop amplitudes for gluon
  fusion into two photons}},  {\em JHEP} {\bf 0109} (2001) 037,
  [\href{http://xxx.lanl.gov/abs/hep-ph/0109078}{{\tt hep-ph/0109078}}].

\bibitem{Bern:2001dg}
Z.~Bern, A.~De~Freitas, L.~J. Dixon, A.~Ghinculov, and H.~Wong, {\it {QCD and
  QED corrections to light by light scattering}},  {\em JHEP} {\bf 0111} (2001)
  031, [\href{http://xxx.lanl.gov/abs/hep-ph/0109079}{{\tt hep-ph/0109079}}].

\bibitem{Bern:2002tk}
Z.~Bern, A.~De~Freitas, and L.~J. Dixon, {\it {Two loop helicity amplitudes for
  gluon-gluon scattering in QCD and supersymmetric Yang-Mills theory}},  {\em
  JHEP} {\bf 0203} (2002) 018,
  [\href{http://xxx.lanl.gov/abs/hep-ph/0201161}{{\tt hep-ph/0201161}}].

\bibitem{Bern:2003ck}
Z.~Bern, A.~De~Freitas, and L.~J. Dixon, {\it {Two loop helicity amplitudes for
  quark gluon scattering in QCD and gluino gluon scattering in supersymmetric
  Yang-Mills theory}},  {\em JHEP} {\bf 0306} (2003) 028,
  [\href{http://xxx.lanl.gov/abs/hep-ph/0304168}{{\tt hep-ph/0304168}}].

\bibitem{Czakon:2013goa}
M.~Czakon, P.~Fiedler, and A.~Mitov, {\it {The total top quark pair production
  cross-section at hadron colliders through O($\alpha_S^4$)}},  {\em
  Phys.Rev.Lett.} {\bf 110} (2013) 252004,
  [\href{http://xxx.lanl.gov/abs/1303.6254}{{\tt arXiv:1303.6254}}].

\bibitem{Boughezal:2013uia}
R.~Boughezal, F.~Caola, K.~Melnikov, F.~Petriello, and M.~Schulze, {\it {Higgs
  boson production in association with a jet at next-to-next-to-leading order
  in perturbative QCD}},  {\em JHEP} {\bf 1306} (2013) 072,
  [\href{http://xxx.lanl.gov/abs/1302.6216}{{\tt arXiv:1302.6216}}].

\bibitem{Ridder:2013mf}
A.~G.-D. Ridder, T.~Gehrmann, E.~Glover, and J.~Pires, {\it {Second order QCD
  corrections to jet production at hadron colliders: the all-gluon
  contribution}},  {\em Phys.Rev.Lett.} {\bf 110} (2013) 162003,
  [\href{http://xxx.lanl.gov/abs/1301.7310}{{\tt arXiv:1301.7310}}].

\bibitem{Gluza:2010ws}
J.~Gluza, K.~Kajda, and D.~A. Kosower, {\it {Towards a Basis for Planar
  Two-Loop Integrals}},  {\em Phys.Rev.} {\bf D83} (2011) 045012,
  [\href{http://xxx.lanl.gov/abs/1009.0472}{{\tt arXiv:1009.0472}}].

\bibitem{Kosower:2011ty}
D.~A. Kosower and K.~J. Larsen, {\it {Maximal Unitarity at Two Loops}},
  \href{http://xxx.lanl.gov/abs/1108.1180}{{\tt arXiv:1108.1180}}.

\bibitem{Larsen:2012sx}
K.~J. Larsen, {\it {Global Poles of the Two-Loop Six-Point N=4 SYM integrand}},
   {\em Phys.Rev.} {\bf D86} (2012) 085032,
  [\href{http://xxx.lanl.gov/abs/1205.0297}{{\tt arXiv:1205.0297}}].

\bibitem{Johansson:2012zv}
H.~Johansson, D.~A. Kosower, and K.~J. Larsen, {\it {Two-Loop Maximal Unitarity
  with External Masses}},  {\em Phys.Rev.} {\bf D87} (2013) 025030,
  [\href{http://xxx.lanl.gov/abs/1208.1754}{{\tt arXiv:1208.1754}}].

\bibitem{CaronHuot:2012ab}
S.~Caron-Huot and K.~J. Larsen, {\it {Uniqueness of two-loop master contours}},
   {\em JHEP} {\bf 1210} (2012) 026,
  [\href{http://xxx.lanl.gov/abs/1205.0801}{{\tt arXiv:1205.0801}}].

\bibitem{Johansson:2013sda}
H.~Johansson, D.~A. Kosower, and K.~J. Larsen, {\it {Maximal Unitarity for the
  Four-Mass Double Box}},  \href{http://xxx.lanl.gov/abs/1308.4632}{{\tt
  arXiv:1308.4632}}.

\bibitem{Sogaard:2013yga}
M.~Sogaard, {\it {Global Residues and Two-Loop Hepta-Cuts}},
  \href{http://xxx.lanl.gov/abs/1306.1496}{{\tt arXiv:1306.1496}}.

\bibitem{Mastrolia:2011pr}
P.~Mastrolia and G.~Ossola, {\it {On the Integrand-Reduction Method for
  Two-Loop Scattering Amplitudes}},  {\em JHEP} {\bf 1111} (2011) 014,
  [\href{http://xxx.lanl.gov/abs/1107.6041}{{\tt arXiv:1107.6041}}].

\bibitem{Badger:2012dp}
S.~Badger, H.~Frellesvig, and Y.~Zhang, {\it {Hepta-Cuts of Two-Loop Scattering
  Amplitudes}},  {\em JHEP} {\bf 1204} (2012) 055,
  [\href{http://xxx.lanl.gov/abs/1202.2019}{{\tt arXiv:1202.2019}}].

\bibitem{Zhang:2012ce}
Y.~Zhang, {\it {Integrand-Level Reduction of Loop Amplitudes by Computational
  Algebraic Geometry Methods}},  {\em JHEP} {\bf 1209} (2012) 042,
  [\href{http://xxx.lanl.gov/abs/1205.5707}{{\tt arXiv:1205.5707}}].

\bibitem{Mastrolia:2012an}
P.~Mastrolia, E.~Mirabella, G.~Ossola, and T.~Peraro, {\it {Scattering
  Amplitudes from Multivariate Polynomial Division}},  {\em Phys.Lett.} {\bf
  B718} (2012) 173--177, [\href{http://xxx.lanl.gov/abs/1205.7087}{{\tt
  arXiv:1205.7087}}].

\bibitem{Mastrolia:2012wf}
P.~Mastrolia, E.~Mirabella, G.~Ossola, and T.~Peraro, {\it {Integrand-Reduction
  for Two-Loop Scattering Amplitudes through Multivariate Polynomial
  Division}},  {\em Phys.Rev.} {\bf D87} (2013) 085026,
  [\href{http://xxx.lanl.gov/abs/1209.4319}{{\tt arXiv:1209.4319}}].

\bibitem{Mastrolia:2013kca}
P.~Mastrolia, E.~Mirabella, G.~Ossola, and T.~Peraro, {\it {Multiloop Integrand
  Reduction for Dimensionally Regulated Amplitudes}},
  \href{http://xxx.lanl.gov/abs/1307.5832}{{\tt arXiv:1307.5832}}.

\bibitem{Badger:2012dv}
S.~Badger, H.~Frellesvig, and Y.~Zhang, {\it {An Integrand Reconstruction
  Method for Three-Loop Amplitudes}},  {\em JHEP} {\bf 1208} (2012) 065,
  [\href{http://xxx.lanl.gov/abs/1207.2976}{{\tt arXiv:1207.2976}}].

\bibitem{Huang:2013kh}
R.~Huang and Y.~Zhang, {\it {On Genera of Curves from High-loop Generalized
  Unitarity Cuts}},  {\em JHEP} {\bf 1304} (2013) 080,
  [\href{http://xxx.lanl.gov/abs/1302.1023}{{\tt arXiv:1302.1023}}].

\bibitem{Cheung:2009dc}
C.~Cheung and D.~O'Connell, {\it {Amplitudes and Spinor-Helicity in Six
  Dimensions}},  {\em JHEP} {\bf 0907} (2009) 075,
  [\href{http://xxx.lanl.gov/abs/0902.0981}{{\tt arXiv:0902.0981}}].

\bibitem{Bern:2010qa}
Z.~Bern, J.~J. Carrasco, T.~Dennen, Y.-t. Huang, and H.~Ita, {\it {Generalized
  Unitarity and Six-Dimensional Helicity}},  {\em Phys.Rev.} {\bf D83} (2011)
  085022, [\href{http://xxx.lanl.gov/abs/1010.0494}{{\tt arXiv:1010.0494}}].

\bibitem{Davies:2011vt}
S.~Davies, {\it {One-Loop QCD and Higgs to Partons Processes Using
  Six-Dimensional Helicity and Generalized Unitarity}},  {\em Phys.Rev.} {\bf
  D84} (2011) 094016, [\href{http://xxx.lanl.gov/abs/1108.0398}{{\tt
  arXiv:1108.0398}}].

\bibitem{Hodges:2009hk}
A.~Hodges, {\it {Eliminating spurious poles from gauge-theoretic amplitudes}},
  \href{http://xxx.lanl.gov/abs/0905.1473}{{\tt arXiv:0905.1473}}.

\bibitem{MR2290010}
D.~Cox, J.~Little, and D.~O'Shea, {\em Ideals, varieties, and algorithms}.
\newblock Undergraduate Texts in Mathematics. Springer, New York, third~ed.,
  2007.
\newblock An introduction to computational algebraic geometry and commutative
  algebra.

\bibitem{MR0463157}
R.~Hartshorne, {\em Algebraic geometry}.
\newblock Springer-Verlag, New York, 1977.
\newblock Graduate Texts in Mathematics, No. 52.

\bibitem{Bern:2002zk}
Z.~Bern, A.~De~Freitas, L.~J. Dixon, and H.~Wong, {\it {Supersymmetric
  regularization, two loop QCD amplitudes and coupling shifts}},  {\em
  Phys.Rev.} {\bf D66} (2002) 085002,
  [\href{http://xxx.lanl.gov/abs/hep-ph/0202271}{{\tt hep-ph/0202271}}].

\bibitem{Bern:1995db}
Z.~Bern and A.~Morgan, {\it {Massive loop amplitudes from unitarity}},  {\em
  Nucl.Phys.} {\bf B467} (1996) 479--509,
  [\href{http://xxx.lanl.gov/abs/hep-ph/9511336}{{\tt hep-ph/9511336}}].

\bibitem{Bern:1996ja}
Z.~Bern, L.~J. Dixon, D.~C. Dunbar, and D.~A. Kosower, {\it {One loop selfdual
  and N=4 superYang-Mills}},  {\em Phys.Lett.} {\bf B394} (1997) 105--115,
  [\href{http://xxx.lanl.gov/abs/hep-th/9611127}{{\tt hep-th/9611127}}].

\bibitem{Gluza:2007rt}
J.~Gluza, K.~Kajda, and T.~Riemann, {\it {AMBRE: A Mathematica package for the
  construction of Mellin-Barnes representations for Feynman integrals}},  {\em
  Comput.Phys.Commun.} {\bf 177} (2007) 879--893,
  [\href{http://xxx.lanl.gov/abs/0704.2423}{{\tt arXiv:0704.2423}}].

\bibitem{Czakon:2005rk}
M.~Czakon, {\it {Automatized analytic continuation of Mellin-Barnes
  integrals}},  {\em Comput.Phys.Commun.} {\bf 175} (2006) 559--571,
  [\href{http://xxx.lanl.gov/abs/hep-ph/0511200}{{\tt hep-ph/0511200}}].

\bibitem{Smirnov:2009up}
A.~Smirnov and V.~Smirnov, {\it {On the Resolution of Singularities of Multiple
  Mellin-Barnes Integrals}},  {\em Eur.Phys.J.} {\bf C62} (2009) 445--449,
  [\href{http://xxx.lanl.gov/abs/0901.0386}{{\tt arXiv:0901.0386}}].

\bibitem{Smirnov:2008py}
A.~Smirnov and M.~Tentyukov, {\it {Feynman Integral Evaluation by a Sector
  decomposiTion Approach (FIESTA)}},  {\em Comput.Phys.Commun.} {\bf 180}
  (2009) 735--746, [\href{http://xxx.lanl.gov/abs/0807.4129}{{\tt
  arXiv:0807.4129}}].

\bibitem{Smirnov:2009pb}
A.~Smirnov, V.~Smirnov, and M.~Tentyukov, {\it {FIESTA 2: Parallelizeable
  multiloop numerical calculations}},  {\em Comput.Phys.Commun.} {\bf 182}
  (2011) 790--803, [\href{http://xxx.lanl.gov/abs/0912.0158}{{\tt
  arXiv:0912.0158}}].

\bibitem{Bern:2006vw}
Z.~Bern, M.~Czakon, D.~Kosower, R.~Roiban, and V.~Smirnov, {\it {Two-loop
  iteration of five-point N=4 super-Yang-Mills amplitudes}},  {\em
  Phys.Rev.Lett.} {\bf 97} (2006) 181601,
  [\href{http://xxx.lanl.gov/abs/hep-th/0604074}{{\tt hep-th/0604074}}].

\bibitem{Carrasco:2011mn}
J.~J. Carrasco and H.~Johansson, {\it {Five-Point Amplitudes in N=4
  Super-Yang-Mills Theory and N=8 Supergravity}},  {\em Phys.Rev.} {\bf D85}
  (2012) 025006, [\href{http://xxx.lanl.gov/abs/1106.4711}{{\tt
  arXiv:1106.4711}}].

\bibitem{Bern:2008qj}
Z.~Bern, J.~Carrasco, and H.~Johansson, {\it {New Relations for Gauge-Theory
  Amplitudes}},  {\em Phys.Rev.} {\bf D78} (2008) 085011,
  [\href{http://xxx.lanl.gov/abs/0805.3993}{{\tt arXiv:0805.3993}}].

\bibitem{Bern:2013yya}
Z.~Bern, S.~Davies, T.~Dennen, Y.-t. Huang, and J.~Nohle, {\it
  {Color-Kinematics Duality for Pure Yang-Mills and Gravity at One and Two
  Loops}},  \href{http://xxx.lanl.gov/abs/1303.6605}{{\tt arXiv:1303.6605}}.

\bibitem{Dixon:1996wi}
L.~J. Dixon, {\it {Calculating scattering amplitudes efficiently}},
  \href{http://xxx.lanl.gov/abs/hep-ph/9601359}{{\tt hep-ph/9601359}}.

\end{thebibliography}

\providecommand{\href}[2]{#2}\begingroup\raggedright\endgroup

\end{document}